\documentclass[nofootinbib, aps, prd,twocolumn,superscriptaddress,longbibliography]{revtex4-1}

\usepackage{amsmath}
\usepackage{amsfonts}
\usepackage[colorlinks=true]{hyperref}
\usepackage{hyperref}
\usepackage{xcolor}

\usepackage{booktabs}

\usepackage{aas_macros}

\usepackage{caption}
\usepackage{graphicx}
\usepackage{subcaption}

\usepackage{bm}

\def\beq{\begin{equation}}
\def\eeq{\end{equation}}

\def\be{\begin{equation}}
\def\ee{\end{equation}}

\def\bea{\begin{eqnarray}}
\def\eea{\end{eqnarray}}

\def\L{\Lambda}

\newcommand{\om}[1]{\mathcal{O}(c^{- #1})}

\newcommand{\<}{\begin{equation}}
\newcommand{\?}{\end{equation}}

\AtBeginDocument{%
    \newwrite\bibnotes
    \def\bibnotesext{Notes.bib}
    \immediate\openout\bibnotes=\jobname\bibnotesext
    \immediate\write\bibnotes{@CONTROL{REVTEX41Control}}
    \immediate\write\bibnotes{@CONTROL{%
    apsrev41Control,author="08",editor="1",pages="0",title="0",year="1"}}
     \if@filesw
     \immediate\write\@auxout{\string\citation{apsrev41Control}}%
    \fi
}%

\begin{document}

\title{Shortcomings of Shapiro delay-based tests of the equivalence principle on cosmological scales}

\author{Olivier Minazzoli}
\affiliation{Centre Scientifique de Monaco, 8 Quai Antoine 1er, 98000, Monaco\\
Artemis, Universit\'e C\^ote d'Azur, CNRS, Observatoire C\^ote d'Azur, BP4229, 06304, Nice Cedex 4, France}

\author{Nathan~K.~Johnson-McDaniel}
\affiliation{Department of Applied Mathematics and Theoretical Physics, Centre for Mathematical Sciences, University of Cambridge, Wilberforce Road, Cambridge,  CB3 0WA, United Kingdom}

\author{Mairi Sakellariadou}
\affiliation{Theoretical Particle Physics and Cosmology Group, Physics Department, King's College London, University of London, Strand, London WC2R 2LS, United Kingdom}

\begin{abstract}
The ``Shapiro delay'' experienced by an astronomical messenger traveling through a gravitational field has been used to place constraints on possible deviations from the equivalence principle. The standard Shapiro delay used to obtain these constraints is not itself an observable in general relativity, but is rather obtained by comparing with a fiducial Euclidean distance. There is not a mapping between the constraints obtained in this manner and alternative theories that exhibit equivalence principle violations. However, even assuming that the comparison with the fiducial Euclidean distance is carried out in a way that is useful for some class of alternative theories, we show that the standard calculation of these constraints cannot be applied on cosmological scales, as is often done. Specifically, we find that the Shapiro delay computed in the standard way (taking the Newtonian potential to vanish at infinity) diverges as one includes many remote sources. We use an infinite homogeneous lattice model to illustrate this divergence, and also show how the divergence can be cured by using Fermi coordinates associated with an observer. With this correction, one finds that the Shapiro delay is no longer monotonic with the number of sources. Thus, one cannot compute a conservative lower bound on the Shapiro delay using a subset of the sources of the gravitational field without further assumptions and/or observational input. As an illustration, we compute the Shapiro delay by applying the Fermi coordinate expression to two catalogs of galaxy clusters, illustrating the dependence of the result on the completeness of the catalogue and the mass estimates.

\end{abstract}

\maketitle

\section{Introduction}
\label{sec:intro}

In the standard model of fundamental physics, different types of electromagnetic and gravitational waves all travel on null geodesics in the geometric optics limit (see, e.g.,~\cite{Isaacson:1967zz}, who demonstrates this for gravitational waves).\footnote{It has been shown that the propagation of gravitational waves does not follow the laws of geometric optics for lenses with masses less than approximately $10^{5} M_{\odot}(f / \mathrm{Hz})^{-1}$, where $f$ is the gravitational wave frequency \cite{takahashi:2017aj}. However, this is not relevant for the cases considered here.} In other words, they may be different either with respect to their respective energies (e.g., gamma versus radio electromagnetic waves), or to the fundamental field they are associated with (e.g., gravitational versus electromagnetic waves), or other properties (e.g., polarization), and still have trajectories that follow from the same null-geodesic equation. Tests of this statement can also be extended to uncharged massive objects with the same mass, e.g., matter versus antimatter, which will follow the same timelike geodesics. This property takes its root from the Einstein equivalence principle that led to general relativity minimally coupled to the standard model of particle physics. Therefore, being able to test the universality of the geodesic equation is a test of the equivalence principle, and any deviation from it would be an indicator of physics beyond the current standard model of physics.
 
As a consequence, there is a large collection of papers dealing with this theme~\cite{1999BASI...27..627S,Desai:2008vj,kahya:2011pl,gao:2015aj,desai:2016mp,wei:2015pl,li:2016ap,nusser:2016ap,wei:2016ap,wu:2016pr,kahya:2016ph,wang:2016pl,tingay:2016ap,wei:2016jc,liu:2017pl,luo:2016jh,sang:2016mn,yang:2016pr,zhang:2017ap,desai:2018ep,Wu:2017yjl,yang:2017mn,yu:2018ap,wei:2017jc,wang:2017ap,boran:2018pr,shoemaker:2018pr,bertolami:2018pd,leung:2018aj,Boran:2018ypz,Laha:2018hsh,wei:2019jh,wei:2019pr,Xing:2019geq,Yao:2019cku}, starting with Longo~\cite{longo:1988pl}, Krauss and Tremaine~\cite{krauss:1988pl}, LoSecco~\cite{LoSecco:1988jg}, and Pakvasa~\emph{et al.}~\cite{pakvasa:1989pr}, although three of these papers actually compared objects following null and non-null geodesics (electromagnetic waves and neutrinos or antineutrinos); LoSecco compares neutrinos and antineutrinos.\footnote{However, at the time those papers were written, it was not known whether neutrinos had a mass, though~\cite{Bose:1988wi} showed that the neutrinos' mass has a negligible effect on these constraints.} All of these papers consider possible differences in the Shapiro delay~\cite{Shapiro:1964uw} experienced by messengers with different properties.\footnote{Note that the ``Shapiro time delay'' is also known as the ``gravitational time delay'' in the lensing community \cite{blandford:1986aj,wambsganss:1998lr}.}
 
The most recent surge of publications on this theme has followed the quasi-coincident detections of gravitational waves (GW170817) and a short-duration gamma-ray burst (GRB 170817A), from which many constraints on the universality of the null-geodesic equation have been derived in the literature, starting from the one of the LIGO, Virgo, Fermi GBM, and INTEGRAL collaborations~\cite{abbott:2017aj}. 

Although there exist various small differences between these analyses, the vast majority of them---starting with Longo~\cite{longo:1988pl}---assume that the metric perturbation is null at infinity and then simply apply the usual parameterized post-Newtonian Shapiro delay equation:
\be
\label{eq:gen_usual_shapiro}
\delta T = - \frac{1+\gamma}{c^3} \int_{\bm r_E}^{\bm r_O} U(\bm{r}(l)) dl +\om{4},
\ee
where $\bm r_E$ and $\bm r_O$ denote emission and observation positions, respectively, $U(\bm{r})$ is the Newtonian gravitational potential, and the integral is computed along the trajectory. $\gamma$ is the parameter appearing in the space-space component of the $c^{-2}$ parametrized post-Newtonian metric. In the parametrized post-Newtonian framework, $\gamma$ will be the same for different messengers, since it is a property of the metric. Nevertheless, for the tests of the equivalence principle we are considering, the standard approach is to assign different values of $\gamma$ to different messengers, though this is not derived from any specific alternative theory or fundamental principle---see the discussion in Sec.~\ref{sec:equiv_principle_bounds}.

Equation (\ref{eq:gen_usual_shapiro}) is gauge dependent, so employing it to constrain deviations from the equivalence principle involves some assumptions, which are usually tacit and are discussed further below. Additionally, it implicitly assumes that the trajectory is short enough that one can treat the region of spacetime containing it as Minkowski plus a linear perturbation, to a good approximation. We will see that this assumption is well justified for sources like GW170817, but not for sources with redshifts $z \gtrsim 1$, like, for instance, the gamma-ray bursts at $z=1.5$, $z = 2.2$, $z = 2.6$, or even $z=11.97$ considered in~\cite{wang:2016pl},~\cite{Wu:2017yjl},~\cite{sang:2016mn}, and~\cite{yu:2018ap}, respectively. While Nusser~\cite{nusser:2016ap} gives a formulation of the constraint that is applicable to more distant sources, the previously cited papers use the standard formulation in Eq.~\eqref{eq:gen_usual_shapiro}.

If one assumes Keplerian potentials (which is a good approximation for all sources that are sufficiently far from the line of sight), the previous equation reduces to \cite{teyssandier2008cq}
\be \label{eq:Shapiro_usual}
\delta T =(1+\gamma) \sum_P  \frac{GM_P}{c^3}\left[\ln\left(\frac{r_{P}+R_{PE}+R_{EO}}{r_{P}+R_{PE}-R_{EO}} \right) \right]+\om{4},
\ee
where $r_{P} := \|\vec x_{P}\|$ and $R_{XY} := \|\vec x_Y - \vec x_X\|$ ($\|\cdot\|$ denotes the Euclidean distance).\footnote{Note that this equation can be given in various equivalent forms by re-arranging the terms in the parentheses in terms of other geometrical quantities. See for instance Eq.~(1) in \cite{longo:1988pl}.} In this situation, one can check that the Shapiro delay is monotonic with respect to the number of sources. As a consequence, one is allowed to consider only a subset of the sources in order to be able to estimate a conservative minimum value of the Shapiro delay, which is necessary in order to give a conservative limit on the violation of the equivalence of the null-geodesic equation.

However, as mentioned above, Eq.~\eqref{eq:gen_usual_shapiro} [and thus Eq.~\eqref{eq:Shapiro_usual}] is gauge dependent, in that it compares the propagation time in the curved spacetime to that of the background Minkowski spacetime in a particular set of coordinates. If one changes the coordinates, then one can obtain both positive and negative values of the time delay, as illustrated in Gao and Wald~\cite{gao:2000cq}. Thus, applications of this expression to equivalence principle constraints tacitly assume that the coordinates used to obtain it are somehow preferred in the context of the equivalence-principle violating theories being tested, as discussed further in Sec.~\ref{sec:basics}.

Another issue with Eq.~(\ref{eq:Shapiro_usual}) is that on cosmological scales---and even if the universe was flat and static---there actually are many distant sources (all the way to infinity, at least in the static universe case). Therefore, the assumption that the gravitational potential is null at infinity is at best an approximation. In what follows, we will show that it is actually an inappropriate assumption on cosmological scales, because in some situations it can lead to an unphysical divergence of the Shapiro delay with the number of the gravitational sources contributing to the delay.\footnote{The divergence already arises at the level of the metric perturbation.} To demonstrate this divergence, we will use an infinite homogeneous lattice toy model, because it allows one to quantify the divergence arithmetically---and because at the same time, it also has been shown to be a good model of the cosmological metric in~\cite{sanghai:2016}. 

What we find is that assuming a null potential at infinity corresponds to imposing an unsuitable choice of gauge (related to an ambiguous choice of the coordinate time), and that the unphysical infinite quantities disappear as soon as one use an appropriate gauge (with a coordinate time related for instance to the proper time of an observer). Nevertheless, we also find that a corrected Shapiro delay equation is no longer monotonic with the number of considered sources, which implies that one cannot simply use a subset of the sources in order to estimate a conservative minimum of the Shapiro delay, which in turns imply that one cannot get a conservative estimate of the test of the universality of the null-geodesic equation, unless the trajectory lies in a region of spacetime where one has measurements of the gravitational field.

Overall, the goal of the present study is to show that, even with the usual tacit assumptions made in constructing the test, it is not possible to obtain a conservative bound on violations of the equivalence principle in cosmological situations with the standard method.

As a consequence, we think that most (if not all) of the constraints on violations of the equivalence principle from propagation over cosmological distances given in the literature so far should be taken with a great deal of caution.\\

In Sec.~\ref{sec:basics}, we discuss the basics of the Shapiro delay constraints on the equivalence principle considered in the literature, including the tacit assumptions underlying them.
In Sec.~\ref{sec:prelim}, we present a preliminary discussion on the relevance of considering a perturbed Minkowski spacetime on cosmological scales. In Sec.~\ref{sec:lattice}, we use a homogeneous lattice universe in order to arithmetically show that the gauge conditions that are usually implicitly used in the literature lead to a divergence of the Newtonian potential, and we give a cure to this nonphysical divergence. In Sec.~\ref{sec:flat_universe}, using a fiducial gauge that is tied to an observer's proper time, we derive an analytical expression of the Shapiro delay in an infinite inhomogeneous flat Keplerian universe in order to show that the Shapiro delay is no longer monotonic with the number of considered sources in general, contrary to the conventional wisdom. In Sec.~\ref{sec:illustration}, we apply our Shapiro delay expression to two catalogs of galaxy clusters in order to further show the behavior of the Shapiro delay with realistic distributions of matter. We give some concluding remarks in Sec.~\ref{sec:conc}. Additionally, we compute the affine distance in a post-Newtonian metric in Appendix~\ref{app:affine} and give a convergence proof for some lattice sums we consider in the Appendix~\ref{app:convergence}.

\section{Basics of Shapiro delay constraints on the equivalence principle}
\label{sec:basics}

\subsection{Definition of the Shapiro delay}

Shapiro delay-based constraints on the equivalence principle only consider a portion of the full propagation time between two spacetime points---the full propagation time is also known as the one-way propagation time.\footnote{In contrast, Solar System observations of the Shapiro delay (discussed in, e.g.,~\cite{lrr-2014-4}) are based on differential measurements of the delay as the path changes, and not on the comparison of the observed delay with a fiducial value.} This one-way propagation time can be obtained unambiguously from an observable in some situations, and therefore is itself observable in those situations. Indeed, if one considers a stationary spacetime, with the observer and source at rest, then the one-way propagation time is just half of the two-way propagation time, where the two-way propagation time would be the proper time measured by an observer between sending a signal to a reflector and receiving the reflection. 

However, even in a case where the one-way propagation time is observable, one still has to define a way of splitting what one refers to as the Shapiro delay from the total propagation time. Such a splitting is, in general, gauge dependent, as discussed for instance in \cite{gao:2000cq}. In asymptotically flat cases, one obtains the usual expression in Eq.~\eqref{eq:gen_usual_shapiro} in the standard parameterized post-Newtonian (PPN) gauge (see Sec.~2.4 in~\cite{will:1985bk}) used to obtain the PPN expression. This expression is the basis of the original tests in Longo~\cite{longo:1988pl} and Krauss and Tremaine~\cite{krauss:1988pl}.

Nevertheless, again assuming asymptotic flatness, one can also obtain the same Shapiro delay expression (in the GR case), up to higher-order PN corrections, by considering the (gauge-invariant) affine distance $d_\text{aff}$ (discussed in, e.g., Sec.~2.4 of~\cite{Perlick:2004tq}). We compute the affine distance for a general post-Newtonian metric in Appendix~\ref{app:affine}, showing that it gives the distance one would na{\"\i}vely compute using the background Minkowski metric in the coordinates in which the post-Newtonian metric is given for an observer at rest with respect to that background metric.
One then obtains the Shapiro delay (plus higher-order PN effects) by subtracting $d_\text{aff}/c$ (where $c$ is the speed of light) from the total one-way propagation delay. This assumes that the metric is known, which allows one to compute the affine distance.
Therefore, at least in asymptotically flat and stationary cases, there is a gauge-invariant definition of the Shapiro delay. 
However, this definition is observer-dependent, because the affine distance depends on the observer 4-velocity.
One notably gets the standard definition for a specific class of observers which are at rest with respect to the fiducial background metric. Also, note that while the affine distance is gauge invariant, it cannot be obtained directly from standard astronomical observations.

However, the affine distance is also the distance one would obtain by starting from the angular distance and correcting for the magnification due to gravitational lensing [see, e.g., Eq.~(42) in~\cite{Perlick:2004tq} for this definition of the magnification]. In any space-time geometry, and for any theory of gravity in which the reciprocity relation holds and the intensity is conserved, the angular distance $d_\text{ang}$ and the luminosity distance $d_\text{lum}$ are related by $d_\text{lum}(z) = (1+z^2) d_\text{ang}(z)$, where $z$ is the redshift~\cite{etherington:1933rm,*etherington:2007sf, ellis:2007vn, ellis:2009fk, ellis:2013fk}. The luminosity distance is an observable for compact binary coalescence observations with gravitational waves as well as for electromagnetic observations of a source with known luminosity (a ``standard candle'').

In practice, for negligible redshift, this means that if one can measure both the luminosity distance $d_\text{lum}$, or the angular distance $d_\text{ang}$, and the one-way propagation time $T$, the Shapiro delay (plus higher-order PN effects) can simply be defined by $\delta T = T - d_\text{lum}/c$, provided that the magnification from gravitational lensing is negligible.\footnote{Note that this definition gives the Shapiro delay in terms of the observer’s proper time.} This could be the case in the Solar System, for instance, if one was in a situation where one can approximate the gravitational field in the Solar System by the stationary gravitational field of the Sun, since the leading contribution to the magnification for small impact parameter gravitational lensing is quadratic in the mass of the lens [see, e.g., Eqs.~(3) and~(9) in~\cite{1988ApJ...332..113A}], while the Shapiro delay is linear in the mass. One could measure the luminosity distance by observing a spacecraft with a known intrinsic transmitter power. However, as mentioned above, this is not how current Shapiro delay constraints are obtained in the Solar System.

Given the strong simplifying assumptions that were necessary to properly define the Shapiro delay, one can expect the analysis to be much more involved in situations in which those simplifying assumptions no longer hold. In particular, if one wants to consider cosmological situations, then it is no longer a good approximation to consider a stationary spacetime: The departures from stationarity are sufficiently large that they cannot be neglected in these calculations, even for the relatively short propagation times appropriate for GW170817/GRB 170817A---they correspond to about $10\%$ of the na{\"\i}ve Shapiro delay, as shown in the next section. While it may be possible to relate the two-way propagation delay to the one-way propagation delay in the cases of interest, this relation would be rather complicated, so we will not pursue this avenue here. 

\subsection{Bounds on equivalence principle violations}
\label{sec:equiv_principle_bounds}

Another thing that is usually assumed in the literature is that the propagation time of waves of different nature would simply be related to Eq.~(\ref{eq:gen_usual_shapiro}) but with a different parameter $\gamma$---which by definition would be a violation of the equivalence principle. This should be seen as a convenient way to infer quantitatively how close the Shapiro time delays must be for waves of different nature, rather than being a parametrization that takes its root from a fundamental theory.

Indeed, as far as we are aware, the method of defining the Shapiro delay to use in equivalence principle tests discussed in the previous subsection is not derived from, or even inspired by, any alternative theories.\footnote{There is some motivation for this sort of expression in Coley and Tremaine~\cite{coley:1988pr}, considering propagation governed by different connections, though it seems tailored to produce the standard Shapiro delay expression. There are also suggestions that this expression can be used to test dark matter emulators~\cite{kahya:2007pl}.} The same seems to be true of the expression used by Nusser~\cite{nusser:2016ap} in the cosmological case, where the Shapiro delay is defined with respect to the background Friedmann-Lema{\^\i}tre-Robertson-Walker (FLRW) cosmology. However, since the standard Shapiro delay expression is used in many studies, for the purposes of this study we will assume that it gives a useful measure of equivalence principle violations in at least some alternative theories (e.g., ones in which there is a preferred coordinate system).\\

Again, the goal of the present study is to show that, even with the strong assumptions listed in this section, it is not possible to obtain a conservative bound on violations of the equivalence principle in cosmological situations---at least not with the standard method.

\section{Preliminaries}
\label{sec:prelim}

In this section, in order to match with the notation in the literature, we set $G = c = 1$. 

In what follows, we show that one can treat the cosmological spacetime as Minkowski plus a linear perturbation for sufficiently nearby sources (including GW170817), and also give some simple arguments for why the standard Shapiro delay calculations are not trustworthy on cosmological scales.

Since the Shapiro delay is usually considered in the post-Newtonian framework, where one is perturbing around a Minkowski background, we first want to write the FLRW background as Minkowski plus a linear perturbation, which one can do for a sufficiently small patch of spacetime around any observer. In order for the approximation of a linear perturbation to be good, the patch has to have dimensions much less than the Hubble radius $L_H := H_0^{-1} \simeq 4.4 \text{ Gpc}$, here taking the current value of the Hubble parameter $H(t)$ to be $H_0 = 68 \text{ km s}^{-1} \text{ Mpc}^{-1}$; cf.~\cite{Aghanim:2018eyx}. Specifically, we are interested in a patch containing an observer on the Earth and the source of the radiation under investigation and expand the metric in powers of $H\|\vec{x}\|$, where $\vec{x}$ is the displacement vector from the Earth.\footnote{Here we use ``the Earth'' as a shorthand for a geodesic of the FLRW metric close to the actual position of the Earth.} This combination is the natural dimensionless expansion parameter for this situation. In fact, we write the FLRW metric in Fermi normal coordinates, where we have [e.g., Eq.~(4) of~\cite{Nicolis:2008in}]
\be\label{eq:FLRW_Fermi}
\begin{split}
ds^2_\text{FLRW} &= -\{1 - [\dot{H}(t) + H^2(t)]\|\vec{x}\|^2\}dt^2\\ 
&\quad + [1 - H^2(t)\|\vec{x}\|^2/2]\|d\vec{x}\|^2 + O(H^4\|\vec{x}\|^4).
\end{split}
\ee
(See~\cite{Holz:1997ic} for related discussion.) Here overdots denote derivatives with respect to cosmic time, so the Friedmann equations for a spatially flat cosmology give $H^2 = (8\pi\rho + \Lambda)/3$ and $\dot{H} + H^2 = (-4\pi\rho + \Lambda)/3$, where $\rho$ is the average energy density and $\Lambda$ is the cosmological constant.

We see that the neglected term is of order $10^{-8}$ for the distance of $\sim 40$~Mpc appropriate for GW170817 (see, e.g.,~\cite{Abbott:2018wiz}) and is thus completely negligible compared to the first-order perturbation, which is of the order of $10^{4}$ larger. However, this term is of order unity for distances around $L_H \simeq 4.4$~Gpc or greater, i.e., $z\gtrsim 1$, so one cannot treat the metric as Minkowski plus a first-order perturbation in those cases. We shall not consider such cases further here.

We now want to consider the case in which we have a perturbation to FLRW from, e.g., a galaxy or galaxy cluster. As a simple model, we will use the McVittie spacetime~\cite{McVittie:1933zz}, given in modern notation in, e.g., Eq.~(20) of~\cite{Shaw:2005gt}, which describes a spherically symmetric mass embedded in an expanding universe. We will only consider this spacetime well away from the singularity at the horizon, so that pathology is not a concern. For the case that gives a spatially flat FLRW metric in the limit where the embedded mass is zero, the McVittie metric in isotropic coordinates takes the form
\begin{subequations}
\begin{gather}
ds^2_\text{McV} = -\left[\frac{1 - \mu(\tau,\rho)}{1 + \mu(\tau,\rho)}\right]^2d\tau^2 + a^2(\tau)\left[1 + \mu(\tau,\rho)\right]^4\|d\vec{y}\|^2,\\
\mu(\tau,\rho) := \frac{m}{2\rho a(\tau)},
\end{gather}
\end{subequations}
where $\tau$ and $\vec{y}$ are cosmic time and the comoving spatial coordinates, respectively (using the same notation as in~\cite{Nicolis:2008in}), $\rho := \|\vec{y}\|$, $a(\tau)$ is the scale factor, and $m$ is the mass parameter of the embedded mass, which is equal to the Schwarzschild mass when $a(\tau) = 1$. We now want to express this metric in coordinates similar to the Fermi normal coordinates used for the FLRW metric in Eq.~\eqref{eq:FLRW_Fermi}. If we apply the coordinate transformation given in Eq.~(3) of~\cite{Nicolis:2008in}, we obtain
\be\label{eq:McV_Fermi}
\begin{split}
ds^2_\text{McV} &= -\left\{1 - [\dot{H}(t) + H^2(t)]\|\vec{x}\|^2 - \frac{2m}{\|\vec{x}\|}\right\}dt^2\\ 
&\quad - \frac{8m H(t)}{\|\vec{x}\|}\vec{x}\cdot d\vec{x} \, dt\\
&\quad + \left[1 - \frac{H^2(t)\|\vec{x}\|^2}{2} + \frac{2m}{\|\vec{x}\|}\right]\|d\vec{x}\|^2\\
&\quad + \mathcal{O}\left(H^4\|\vec{x}\|^4,\frac{m^2}{\|\vec{x}\|^2}, mH^2\|\vec{x}\|\right).
\end{split}
\ee
Except for the mixed spatiotemporal terms, this is the same metric one would obtain if one na{\"\i}vely superposed the Fermi normal coordinate linearized FLRW metric [Eq.~\eqref{eq:FLRW_Fermi}] and the standard Newtonian order post-Newtonian metric of a point mass usually used to compute the Shapiro delay. Note, however, that the metric in Eq.~(\ref{eq:McV_Fermi}) no longer is in Fermi coordinates, compared to the metric in Eq.~(\ref{eq:FLRW_Fermi}), because of the perturbing mass terms.

In fact, the spatiotemporal terms are much smaller than the perturbations in the diagonal terms for the example case we consider. Specifically, we consider the same setup used in the equivalence principle constraint from the GW170817/GRB 170817A signals in~\cite{abbott:2017aj}, using the Milky Way's Keplerian potential contribution with a mass $m = 2.5\times 10^{11} M_\odot$ and minimum and maximum distances of $r_0 = 100$~kpc and $r_1 = 26$~Mpc (the $90\%$ credible level lower bound on the distance obtained solely from gravitational waves from the analyses performed at the time of that paper~\cite{abbott:2017pl}). We consider a radial path, for simplicity, and also use the cosmological parameters $H_0 = 68 \text{ km s}^{-1} \text{ Mpc}^{-1}$, $\Omega_\text{m} = 0.31$, and $\Omega_\Lambda = 1 - \Omega_\text{m}$ (cf.\ the TT,TE,EE+lowE+lensing+BAO parameters in Table~2 of~\cite{Aghanim:2018eyx}). The na{\"\i}ve Shapiro delay including the spatiotemporal terms is given (suppressing the remainder) by
\be
\label{eq:McV_shapiro}
\begin{split}
&\quad\int_{r_0}^{r_1}\left[\frac{2m}{r} + \frac{H^2(t(r)) + 2\dot{H}(t(r))}{4}r^2 - 4mH(t(r))\right] dr\\
&= 2m\ln\left(\frac{r_1}{r_0}\right) + \biggl[\frac{(1 - 3\Omega_\text{m})H_0^2}{12}(r_1^3 - r_0^3)\\
& \quad + \frac{3\Omega_\text{m}H_0^3}{8}(r_1^4 - r_0^4)\biggr] - 4mH_0(r_1 - r_0),
\end{split}
\ee
where we have used $H^2(t) + 2\dot{H}(t) = H^2_0 + 2\dot{H}_0 + 2(H_0\dot{H}_0 + \ddot{H}_0)t + \mathcal{O}(H_0^2t^2) = [1 - 3(1 - 2H_0 t)\Omega_\text{m}]H_0^2 + \mathcal{O}(H_0^2t^2)$, noted that $t(r) = -r$, and neglected the contribution from the time dependence to the spatiotemporal terms, since these already give a small contribution. 
The contributions of the three terms in the sum (taking the terms in brackets as a single term here) to the na{\"\i}ve Shapiro delay are $\sim 160$~days, $\sim 19$~years (from individual contributions of $\sim 15$ and $\sim 2$ years), and $\sim -8$~hours, respectively, corresponding to the contribution from the Keplerian potential, the cosmological curvature, and the spatiotemporal terms. We thus see that the spatiotemporal terms are indeed negligible, although note that they contribute negatively to the effect. The small factor of $1 - 3\Omega_\text{m} \simeq 0.07$ is part of why the time dependence (second term in brackets) is a relatively large correction to the time-independent contribution (first term in brackets).

Additionally, the contribution from the Keplerian potential is much smaller than that from the cosmological curvature. Thus, it might seem that one can compute this na{\"\i}ve Shapiro delay using only the cosmological curvature to a good approximation, since it should take into account the contributions from the many distant galaxies that yield the diverging contribution to the Shapiro delay (see Sec.~\ref{sec:diverge}) in the na{\"\i}ve calculation using Eq.~\eqref{eq:Shapiro_usual}. This may in fact be the case, but the Shapiro delay obtained in this manner would not necessarily give a conservative bound, since there will be negative contributions to the metric perturbation, and thus the Shapiro delay, from nearby underdensities. Thus, a more careful calculation is necessary to assess the size of these contributions to the Shapiro delay. This could in principle be substantial, given the maximum size of the Newtonian potential of $\sim 10^{-4}$ mentioned in~\cite{Green:2014aga}, which is an order of magnitude larger than the maximum of the cosmological curvature term in the Fermi coordinate FLRW metric [Eq.~\eqref{eq:FLRW_Fermi}] at the $26$~Mpc distance. It is possible that a statistical argument like in Nusser~\cite{nusser:2016ap} could be used to give constraints on deviations from the cosmological curvature, even in the absence of direct measurements of the gravitational field along the line of sight. 


\section{The infinite homogeneous lattice}
\label{sec:lattice}

In order to show the unrealistic divergence of the metric perturbation with the number of sources, it is convenient to use a model where the perturbation can be computed arithmetically. An infinite homogeneous lattice model allows such computations. Additionally, it has been shown to produce the results of the usual standard model of cosmology with ``small'' back-reaction effects that are due to the discreteness of the perturbations considered \cite{sanghai:2016}. In such a model, the universe is filled with an infinite lattice of cells. All cells possess the same mass at the same relative location, and the metric is given in terms of a post-Newtonian expansion in each cell.\footnote{Recall that the post-Newtonian expansion assumes weak fields and slow motions, as well as a Minkowski background at the scale of the phenomenon considered---here, within a cell.} The evolution of the universe then follows from junction conditions between cells, known as the Israel junction conditions.  Therefore, such a model allows us to compute the Shapiro delay from an infinite set of masses uniformly distributed, in an otherwise realistically evolving universe.

\subsection{Metric within each cell}

Given a cell with a sufficiently small size $L(t)$~\cite{sanghai:2016,sanghai:2015}, the metric can be expanded around Minkowski spacetime and reads:
\beq \label{eq:metric_sanghai}
ds^2 = \left(-1 + h_{00} \right) c^2 dt^2 + \left( \delta_{ij} + h_{ij}  \right) dx^i dx^j + \om{3},
\eeq
where $h_{00}$ and $h_{ij}$ are the $c^{-2}$ perturbations of the metric (Latin letters denote spatial indices), such that \cite{sanghai:2016}
\bea
&&h_{00} \equiv 2 \Phi = 2 \left( \Phi_M +  \Phi_\L \right), \\
&&h_{ij} \equiv 2 \Psi  \delta_{ij} =  2 \left( \Phi_M - \frac{ \Phi_\L}{2} \right) \delta_{ij}.
\eea
This metric is expressed in standard post-Newtonian coordinates, so it takes this simple diagonal form to leading order~\cite{sanghai:2015}.
At leading order in the post-Newtonian approximation, the Einstein equation with a cosmological constant then reduces to
\beq
\triangle \Phi_M = - \frac{4 \pi G}{c^2} \sum_{\vec \beta \in \mathbb{Z}^3} M \delta^{(3)} \left(\vec x - L(t) \vec \beta \right),
\eeq
and
\beq
\triangle \Phi_\L = \L,
\eeq
where $\triangle$ is the flat-space Laplacian and $M$ is the mass in each cell. Let us stress that in the lattice model~\cite{sanghai:2016,sanghai:2015}, one is solving the equations in a single cell, so the appropriate boundary conditions are those at the cell boundary, not at infinity. The solution can be decomposed such that
\beq
\Phi_\L = \frac{\L}{6} r^2,
\eeq
with $r^2 := x^2+y^2+z^2$, and
\bea
\Phi_M = \frac{GM}{r c^2} + \Phi_t (t)+ \delta \Phi_l(t) x^l,
\eea
with the tidal potential~\cite{sanghai:2016,sanghai:2015}
\beq
\Phi_t := \frac{GM}{c^2} \sum_{\vec \beta \in \mathbb{Z}^3_*} \frac{1}{\|\vec x - L(t) \vec \beta \|} + \Phi_0(t),
\eeq
where $\mathbb{Z}^3_* := \mathbb{Z}^3 \setminus \{\vec{0}\}$ and $\Phi_0(t)$ and $\delta \Phi_l(t)$ are gauge-dependent terms.  In what follows, we set the mass at the origin of the lattice equal to 0, in order to consider an observer located at the origin, which simplifies the expressions. We thus have
\bea
\Phi_M = \Phi_t (t)+ \delta \Phi_l(t) x^l,
\eea

\subsection{An unsuitable choice of coordinate time }
\label{sec:diverge}

When considering a problem with a finite number of bodies, the gauge is often restricted to the case that satisfies
\bea \label{eq:usual_gauge}
\lim_{r \rightarrow \infty} \Phi_M  =\Phi_0(t) =0,\\
\lim_{r \rightarrow \infty} \partial_l \Phi_M = \delta \Phi_l(t) =0, \label{eq:usual_gauge2}
\eea
 such that the coordinate time corresponds to the proper time of an ideal observer situated at infinity. Then, if one considers, for instance, Solar System observables, one simply converts this coordinate time to the proper time of an actual observer. However, when there are (non-negligible) sources located all the way up to infinity, the metric perturbation cannot be taken to be null at infinity, so the previous gauge restrictions simply do not make sense when there are sources located all the way up to infinity. The infinite lattice model allows us to illustrate this from a simple arithmetic point of view: $\Phi_t$ is not finite when $\Phi_0 =0$, which is a consequence to the fact that the Epstein zeta function
\beq
\sum_{\vec k \in \mathbb{Z}^3_*}  \frac{1}{\|\vec k\|^n}
\eeq
is only finite for $n >3$. (While this Epstein zeta function can be analytically continued to all of $\mathbb{C}$ with a simple pole at $n = 3$---see, e.g.,~\cite{GlasserZucker}---we will not consider this here, as we are only concerned with cases for which the sums we are considering converge.) Indeed, using the homogeneous property of the lattice model, and defining $\vec \xi := \vec x / L(t)$, one has
\be
\Phi_t(t)-\Phi_0(t) =\frac{GM}{L(t) c^2} \sum_{\vec \beta \in \mathbb{Z}^3_*} \frac{1}{\|\vec \xi - \vec \beta \|}. \label{eq:diverg}
\ee
One can verify that this series is divergent for all locations $\vec x$. As a consequence, the metric perturbation at any given location would not be finite if one naively imposes $\Phi_0(t) = 0$ by hand. As far as we are aware, $\Phi_0(t) = 0$ is always assumed in the literature.

\subsection{Fixing the coordinate time issue}

Interestingly enough, this divergence can be \textit{renormalized} if one chooses instead to restrict the gauge freedom to Fermi coordinates\footnote{There is of course an infinite set of coordinate systems that would allow one to make the calculations without being confronted with infinities.} (i.e., by demanding that the tidal potential $\Phi_t$ and its gradient cancel out at the location associated with an observer). It would mean that one uses a coordinate system that follows the geodesic motion of an observer. Note that this is the usual procedure in the framework of reference frame theory in order to define a proper reference frame (see for instance \cite{soffel:2003}). In this situation, the tidal potential would read
\beq
\label{eq:tidal_pot}
\Phi_t(t) = \frac{GM}{L(t) c^2} I(\vec \xi),
\eeq
with
\be\label{eq:tidal_pot_def}
I(\vec \xi) := \sum_{\vec \beta \in \mathbb{Z}^3_*} \left( \frac{1}{\|\vec \xi - \vec \beta\|} - \frac{1}{ \|\vec \beta\|} \right).
\ee
One can verify that $I(\vec \xi)$ is indeed finite (with the appropriate prescription for the lattice summation, since it is a conditionally convergent series---see Appendix~\ref{app:convergence}) for all $\vec \xi \notin \mathbb{Z}^3_*$, while the two members of the sum lead to divergent series when taken separately. Let us also note that, by construction, $I(\vec 0) =0$.

With the previous gauge restriction, the potential
reduces to
\bea
\Phi_M = \frac{GM}{c^2} \left[ \frac{1}{r}+\frac{I(\vec \xi)}{L(t)}\right]+ \delta \Phi_l(t) x^l,
\eea
where $I(\vec \xi)$ is defined in Eq.~(\ref{eq:tidal_pot_def}).

The gradient of the potential on the other hand reads
\be
\partial_i \Phi_M = - \frac{GM}{L^2(t) c^2} J^i(\vec \xi)+ \delta \Phi_i,
\ee
with
\be\label{eq:tidal_pot_grad_sum}
J^i(\vec \xi) := \sum_{\vec \beta \in \mathbb{Z}^3_*} \frac{\xi^i - \beta^i}{\|\vec \xi - \vec \beta\|^3} .
\ee
One can verify that $J^i(\vec \xi)$ is also finite. Additionally, $J^i(\vec 0)$ is obviously equal to zero thanks to the symmetry of the sum (if the sum is performed in a way that respects this symmetry---see Appendix~\ref{app:convergence}). The last touch in order to select Fermi coordinates is to demand that $\partial_i \Phi_M({\vec x}_O)=0$, where ${\vec x}_O$ is the position of the observer, thus fixing $\delta \Phi_i$. For an observer at the origin, we thus have $\delta \Phi_i = 0$.

\section{The infinite inhomogeneous flat Keplerian universe}
\label{sec:flat_universe}

Now that we have seen arithmetically with the lattice model why one needs to carefully chose a non-ambiguous coordinate time in order to have ``meaningful'' potentials in the metric, we can derive the Shapiro delay more properly. In most of the literature, the universe is taken to be flat and perturbed by either some Keplerian potentials, or by some other types of potentials that are meant to depict dark matter in halos. We are not interested in a detailed study here because our goal is to show that a conservative estimate of the Shapiro delay cannot be obtained with these kind of calculations anyway. Hence, we will limit our study and assume Keplerian potentials. Also, we now consider a parametrized post-Newtonian metric in order to be able to quantify the difference of the Shapiro delays for different messengers the way it is done in the literature---see Sec.~\ref{sec:equiv_principle_bounds}. Let us note that a parametrized post-Newtonian metric is compatible with the lattice model, see \cite{sanghai:2017}. The metric we consider therefore reads
\be
ds^2 = (-1+2 \Phi) c^2 dt^2 + (1+2 \gamma \Phi)\delta_{ij} dx^i dx^j + \om{3}, \label{eq:metric_gamma}
\ee
where
\be
\Phi = \frac{G }{c^2}\sum_P \frac{M_P}{\|\vec x - {\vec x}_P\|} + \Phi_0(t) + \delta \Phi_l(t) x^l.
\ee
We choose our coordinate system such that they are Fermi coordinates associated to the observer:
$\Phi(\vec x= {\vec x}_O=\vec{0})=0$ and $\partial_i \Phi(\vec x= {\vec x}_O=\vec{0})=0$. Therefore, the potential reduces to\footnote{See also Eq.~(28) in \cite{soffel:2003}, or the leading order of Eq.~(4.31) in \cite{klioner:2000pr}.}
\be
\Phi = \sum_P \frac{G M_P}{c^2} \left[\frac{1}{\|\vec x - {\vec x}_P\|} -  \frac{1}{\|{\vec x}_P\| }-  \frac{\vec x \cdot {\vec x}_P}{\|{\vec x}_P\|^3}  \right].
\ee
Note that, while the observer is at rest in the Fermi coordinate system, the source of gravitational and electromagnetic waves is not. However, we assume that the source is at a cosmological distance and emits the messengers being considered over a short time period, such that any motion of the source has a negligible effect on the path the messengers take. Of course, this expression is only valid when $\Phi \ll 1$, as was assumed in the derivation. In particular, the final term can be large for large $\|{\vec x}\|$. This term is associated with the forces acting on the observer in the Newtonian picture, and its contribution to the sum vanishes if the source mass density is isotropic. We have checked that one indeed has $\Phi \ll 1$ in the application to galaxy clusters that we consider.

Assuming that a given wave travels on the null geodesics of the metric (\ref{eq:metric_gamma}), one has $cdt = \left[1+(1+\gamma) \Phi \right] dl + \om{3}$. Using the fact that a null geodesic is a straight line at leading order when one is far from the ``lensing regime''---defined in \cite{ashby:2010cg,linet:2016pr} as the regime for which multiple images can appear---this equation can be analytically solved, such that the propagation time between an emission point ${\vec x}_E$ and the observer is
\begin{widetext}
\be \label{eq:Shapiro_kepl}
T(\vec x_E, \vec x_O) = \frac{R_{EO}}{c}  +\left(1+\gamma \right) \sum_P  \frac{GM_P}{c^3}\left[\ln\left(\frac{r_{P}+R_{PE}+R_{EO}}{r_{P}+R_{PE}-R_{EO}} \right)-  \frac{R_{EO}}{r_{P}} - \frac{1}{2} \left(\frac{R_{EO}}{r_{P}}\right)^2 \cos \theta_{P} \right]+\om{4},
\ee
\end{widetext}
where $r_{P} := \|\vec x_{P}\|$, $R_{XY} := \|\vec x_Y - \vec x_X\|$, and with $\cos \theta_{P} := ({\vec x}_O-{\vec x}_E) \cdot {\hat x}_P/\|{\vec x}_O-{\vec x}_E\|$. Note that, since one uses Fermi coordinates associated with the observer, the propagation time is expressed in terms of the observer proper time---as it should be.

\subsection{Constraints on the equivalence principle}

Now, let us assume that one is able to measure the difference of propagation time between an event and the observer of the two types of waves $X$ and $Y$. Then, according to Eq.~(\ref{eq:Shapiro_kepl}), and with the assumptions about the nature of the equivalence principle violation mentioned in Sec.~\ref{sec:equiv_principle_bounds}, the fractional difference between the two propagation times would read
\begin{widetext}
\be
\label{eq:Shapiro_new}
\Delta T_{XY}=\left(\gamma_Y - \gamma_X \right) \sum_P  \frac{GM_P}{c^3}\left[\ln\left(\frac{r_{P}+R_{PE}+R_{EO}}{r_{P}+R_{PE}-R_{EO}} \right)-  \frac{R_{EO}}{r_{P}} - \frac{1}{2} \left(\frac{R_{EO}}{r_{P}}\right)^2 \cos \theta_{P} \right]+\om{4}.
\ee
\end{widetext}

Because the terms inside the brackets do not all have the same sign, one cannot simply use a subset of the sources in order to get a conservative constraint on the difference $\gamma_Y - \gamma_X$, unlike what is usually assumed in the literature. As a consequence, unless one has an absolute knowledge of the location and the mass of all the sources, or the propagation occurs only in a region in which one has measurements of the gravitational field, one cannot give a conservative constraint on the difference $\gamma_Y - \gamma_X$ from this type of calculation.

\section{Illustration with GW-GRB 170817 and two catalogs of clusters}
\label{sec:illustration}

As an illustration, we apply the analytical equation (\ref{eq:Shapiro_kepl}) to the case of the (quasi-)coincident detection of the gravitational wave GW170817 and the short duration gamma ray burst GRB 170817A.  We use two recent catalogs of galaxy clusters in order to model the distribution of the gravitational sources. 

The first catalog we will consider is based on a \textit{friends-of-friends} finding algorithm \cite{tempel:2016aa}. We will refer to it as \textit{Tempel2016}. The second catalog, on the other hand, calibrates the group finder from a halo occupation model \cite{tully:2015aj}. We will refer to it as \textit{Tully2015}. The former catalog is built on the 2MRS, CF2, and 2M++ survey data comprising nearly $80,000$ galaxies within the local volume of 430 Mpc radius; while the latter is built from a sample of the 2MASS Redshift Survey almost complete to $K_s = 11.75$ over 91\% of the sky, which has about $43,000$ entries, giving a maximum distance of 240~Mpc. In Tully2015, the clusters' mass is either obtained from adjusted intrinsic luminosity and mass to light prescription, or obtained from the virial theorem.  In Tempel2016, the masses are solely obtained from the virial theorem.

The code that derives the following results is freely accessible~\cite{code}. 

\begin{table*}
\caption{\label{tab:cat_Shapiro} The Shapiro delays (with $\gamma = 1$) for GW170817/GRB 170817A using the position and sky location given in the text and computed using the different catalogs considered and either Eq.~(\ref{eq:Shapiro_kepl}) or Eq.~(\ref{eq:Shapiro_usual}), i.e., either the new expression or the usual, incorrect expression (which is equivalent to taking $\Phi_0 =0$ and $\delta \Phi_l =0$). We also give the number of sources used in each calculation.}
\begin{tabular}{|l|c|c|c|}
\hline 
catalog & \# of sources & Shapiro delay with new Eq.~(\ref{eq:Shapiro_kepl}) [yr] & Shapiro delay with old Eq.~(\ref{eq:Shapiro_usual}) [yr] \\
\hline
Tempel2016 & $5,166$ & $+~77$ & $+~22,985$\\
Tully2015 (luminosities) & $25,472$ & $+~65$ & $+~190,502$ \\
Tully2015 (virial) & $1,119$ & $-~20$ & $+~37,282$ \\ \hline
\bottomrule
\end{tabular}
\end{table*}

In Table~\ref{tab:cat_Shapiro} we give the results for the Shapiro delay for GW170817/GRB 170817A using the same parameters as in~\cite{abbott:2017aj} (a distance of $26$~Mpc and the sky location from~\cite{Coulter:2017wya}). For comparison, the value obtained with just the Milky Way's Keplerian potential outside of $100$~kpc (as in~\cite{abbott:2017aj}) is $79$~days. We find that the values obtained with the standard, incorrect expression [Eq.~(\ref{eq:Shapiro_usual})] are more than two or three orders of magnitude larger than the values obtained with the new expression [Eq.~(\ref{eq:Shapiro_kepl})], and even have opposite sign in the Tully2015 virial mass case. 

\subsection{Comparison with cosmology}

It is interesting to compare the estimate of the Shapiro delay when using either the catalogs or the FLRW metric in Fermi coordinates. Indeed, since the FLRW metric assumes a homogeneous mean density, one can expect that an average of the Shapiro delay over the whole sky in the catalogs gives a value with the same order of magnitude with respect to an estimate from the matter contribution alone in an FLRW universe. Note, however, that this comparison cannot be rigorous, as one compares calculations in gauges that use different prescriptions.
Assuming the Friedmann equations given in Sec.~\ref{sec:prelim}, and the line element [Eq.~(\ref{eq:FLRW_Fermi})], the total Shapiro delay reads
\be
\label{eq:T_FLRW}
T(\vec x_E, \vec x_O) =  \frac{(\Omega_\Lambda - 2\Omega_\text{m})H_0^2}{12}(r_E^3 - r_O^3). 
\ee
The contribution from matter alone therefore reads
\be
T_\textrm{matter}(\vec x_E, \vec x_O) = - \frac{\Omega_\text{m}H_0^2}{6}(r_E^3 - r_O^3).
\ee
The contribution from matter alone in an FLRW universe would therefore be about $-152$~years for a source located at 26 Mpc. 

\begin{figure}[h!]
\includegraphics[scale=0.42]{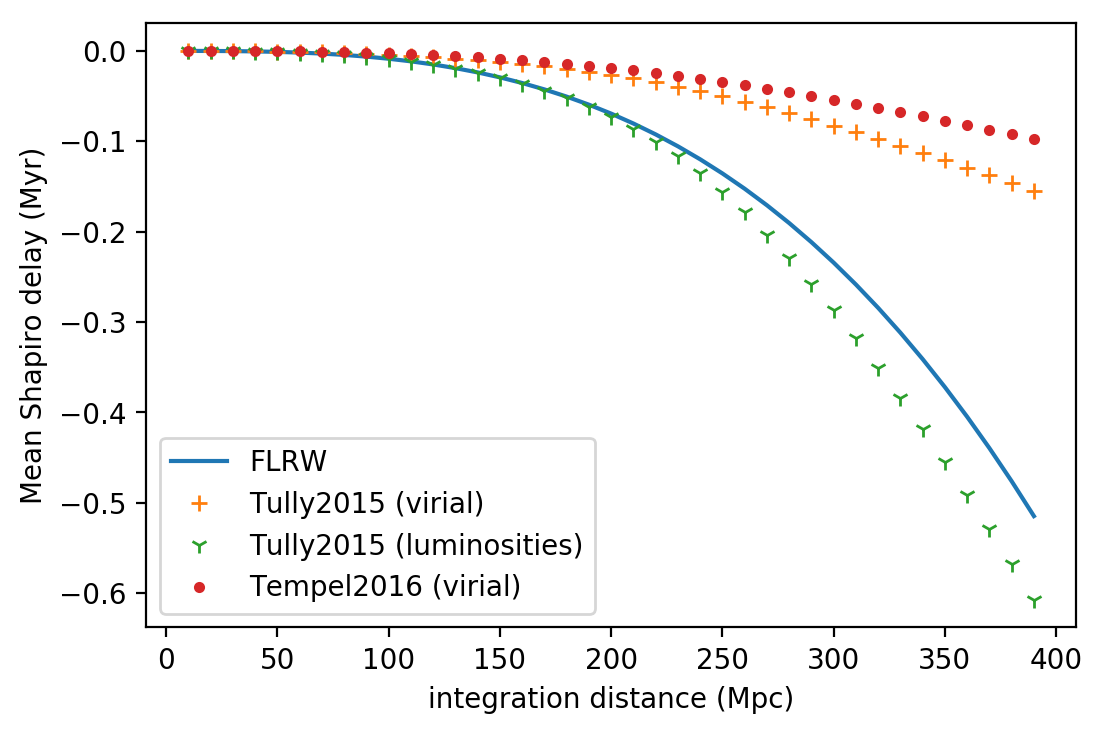}
\caption{Matter contribution to the Shapiro delay in an FLRW universe, as well as an average over the whole sky at a given distance of the Shapiro delay from each catalog, plotted versus the distance to the source.}
\label{img:comparison_FLRW}
\end{figure}

In Fig.~\ref{img:comparison_FLRW}, we plot the evolution of the Shapiro delay with the distance of the source of gravitational waves in an FLRW universe, as well as an average over the whole sky of the Shapiro delay from each catalog. We find that the estimate with the FLRW metric lies in between the estimate with the catalog that determines masses from the luminosity, and the estimates with the catalogs that determine the masses from the virial theorem. This is consistent, given that catalogs that infer masses from the virial theorem cannot estimate all the masses due to a lack of (good enough) data, and therefore they necessarily tend to have an underestimated density of sources. For instance, of the 12,106 sources in Tempel2016, only 5,166 have their mass estimated via the virial theorem.

We therefore conclude that this provides additional evidence that the Shapiro delay in Eq.~(\ref{eq:Shapiro_kepl}) is more appropriate than the one usually found in the literature [Eq.~(\ref{eq:Shapiro_usual})], in the sense that it gives consistent results with the estimation based on a FLRW universe.

\subsection{Discussion on the estimates}

\begin{figure}[h!]
\begin{subfigure}[b]{0.3\textwidth}
\caption{From Tempel2016.}
\includegraphics[scale=0.42]{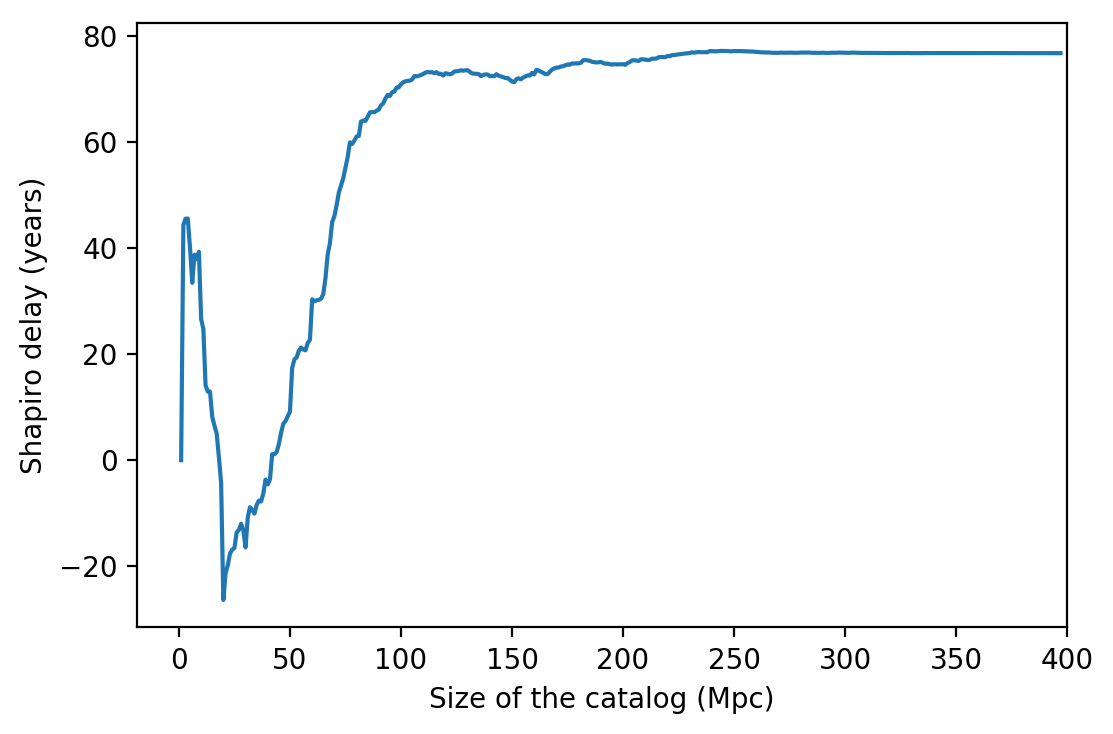}
\end{subfigure}
\begin{subfigure}[b]{0.3\textwidth}
\caption{From Tully2015 with masses inferred from luminosities.}
\includegraphics[scale=0.42]{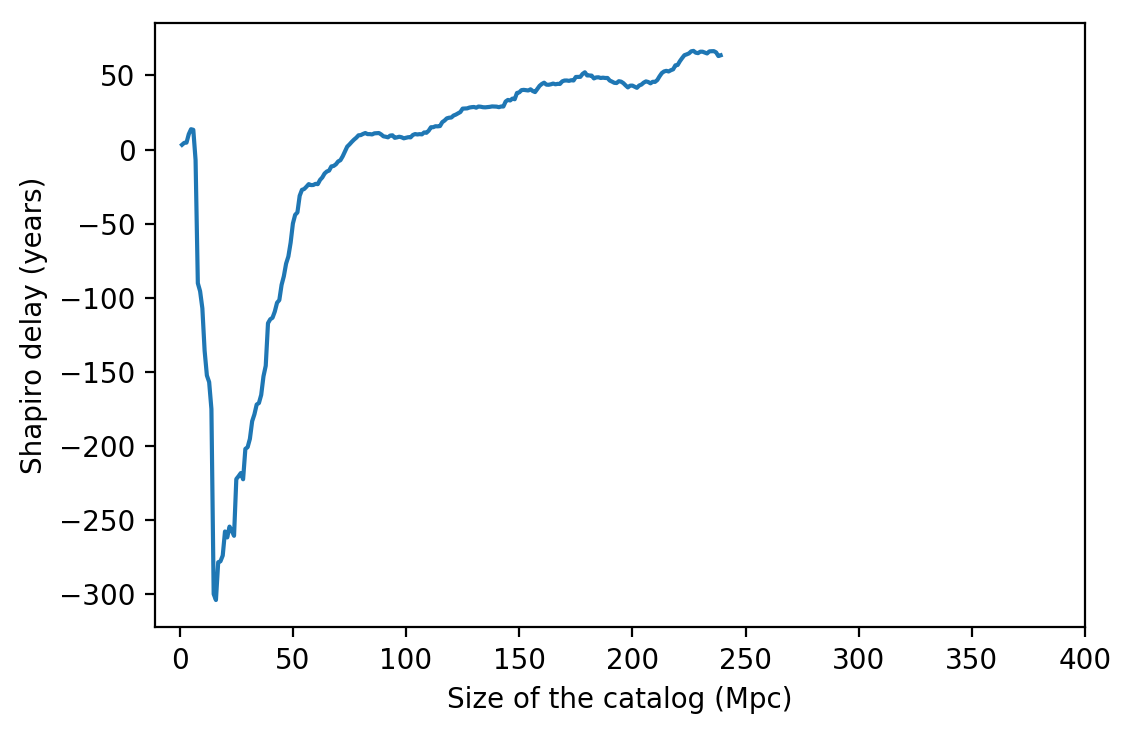}
\end{subfigure}
\begin{subfigure}[b]{0.3\textwidth}
\caption{From Tully2015 with masses inferred from the virial theorem.}
\includegraphics[scale=0.42]{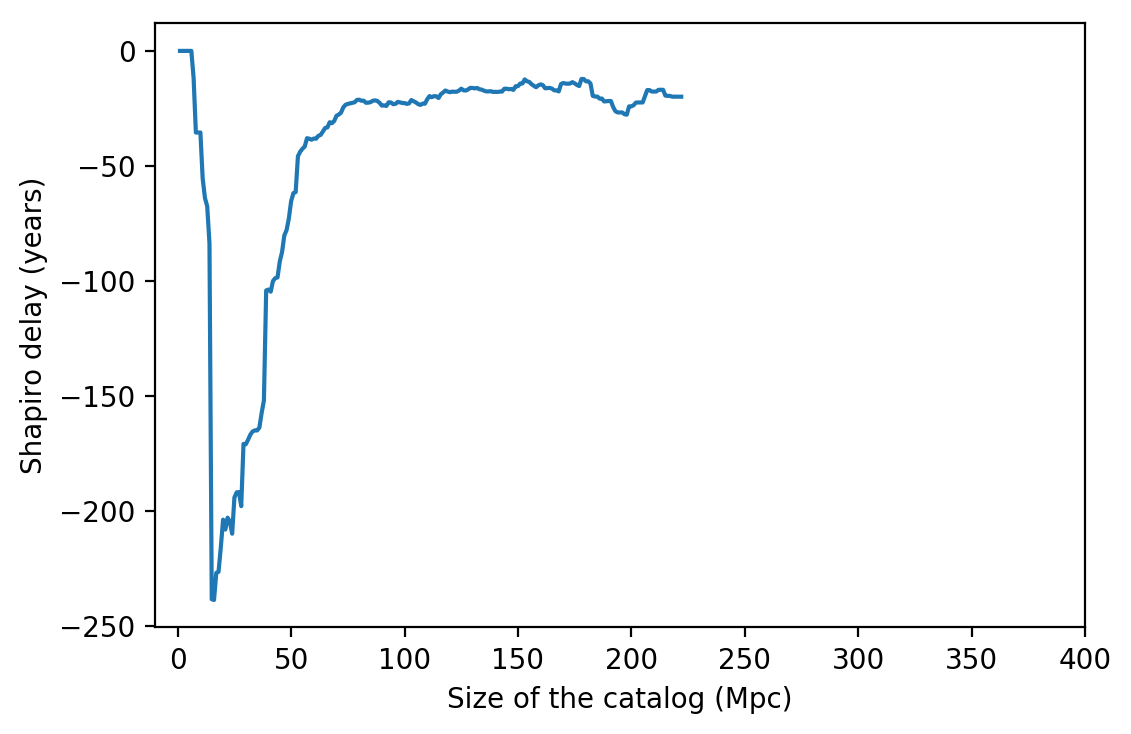}
\end{subfigure}
\caption{Plots of the time delay as one includes more and more remote objects from a catalog. Note that the size of the portion of the universe represented in each catalog is different, with maximum radii of $\sim 400$, $240$, and $220$~Mpc for cases (a), (b), and (c), respectively.}
\label{img:Shapiro_wrt_size}
\end{figure}

It is important to keep in mind that we give those results as an illustration only, and there are not meant to be taken either as rigorous or as conservative estimates of the Shapiro time delay. They are not rigorous because the space-time model that has been used does not depict our universe accurately---in particular, it neglects cosmology (see Secs.~\ref{sec:basics} and~\ref{sec:prelim}). 

But even if one is in a regime and in a theoretical framework that are such that the assumptions that led to Eq.~(\ref{eq:Shapiro_new}) hold, it would not be possible to give conservative constraints on the violation of the equivalence principle with this method from an incomplete set of the gravitational sources anyway (see the previous section). Indeed, adding new sources may result in a decrease of the Shapiro delay, and not necessarily to an increase---unlike what is usually found in the literature. This can be seen with the two catalogs considered here in Fig.~\ref{img:Shapiro_wrt_size}, where we plotted the behavior of the total Shapiro delay (when $\gamma=1$) as one includes more and more remote objects from the catalog. Indeed, one can see that the behavior of the Shapiro delay is not monotonic with the number of sources. This behavior arises because the terms in the bracket of Eq.~(\ref{eq:Shapiro_kepl}) can either be positive or negative depending on the geometrical configuration. 

\begin{figure}[h!]
\begin{subfigure}[b]{0.3\textwidth}
\caption{From Tempel2016.}
\includegraphics[scale=0.42]{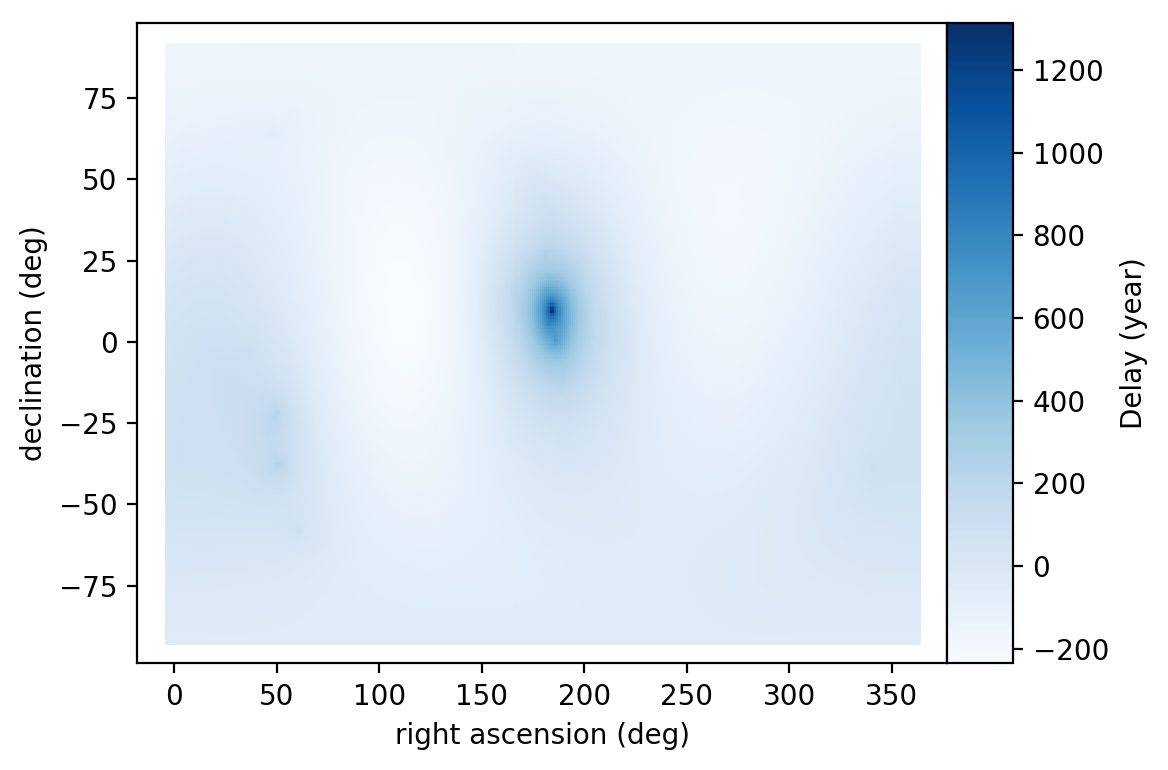}
\end{subfigure}
\begin{subfigure}[b]{0.3\textwidth}
\caption{From Tully2015 with masses inferred from luminosities.}
\includegraphics[scale=0.42]{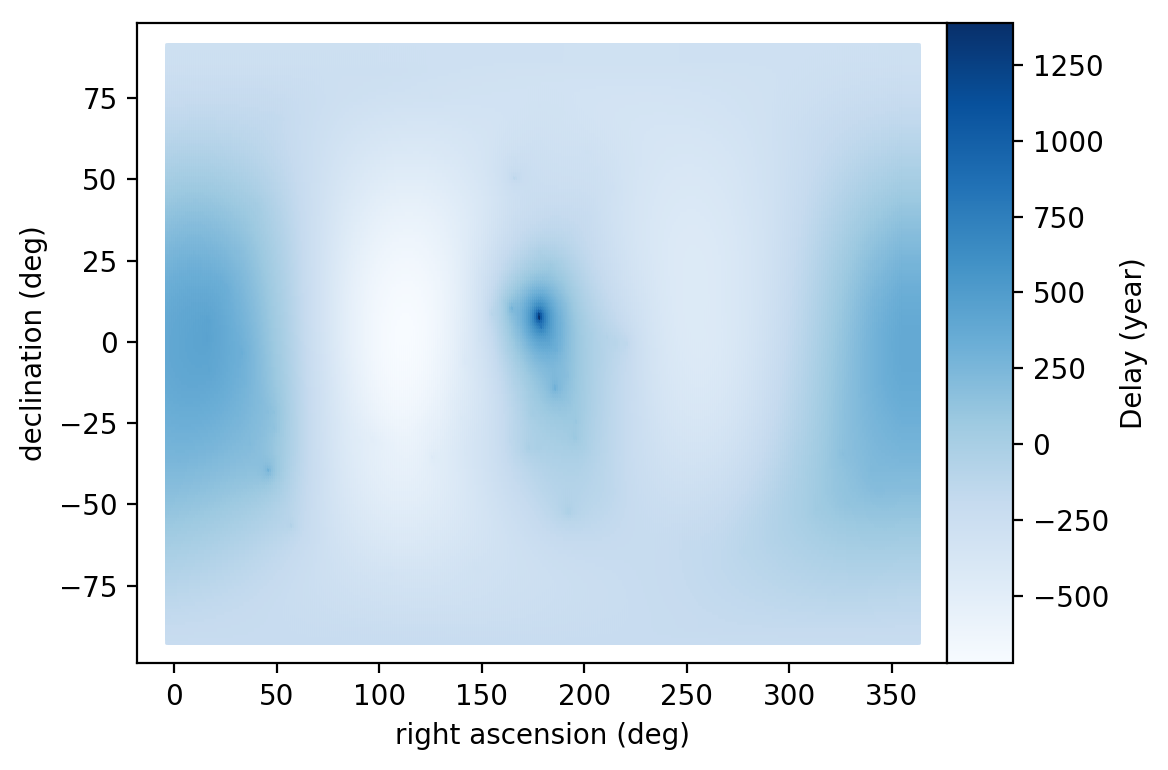}
\end{subfigure}
\begin{subfigure}[b]{0.3\textwidth}
\caption{From Tully2015 with masses inferred from the virial theorem.}
\includegraphics[scale=0.42]{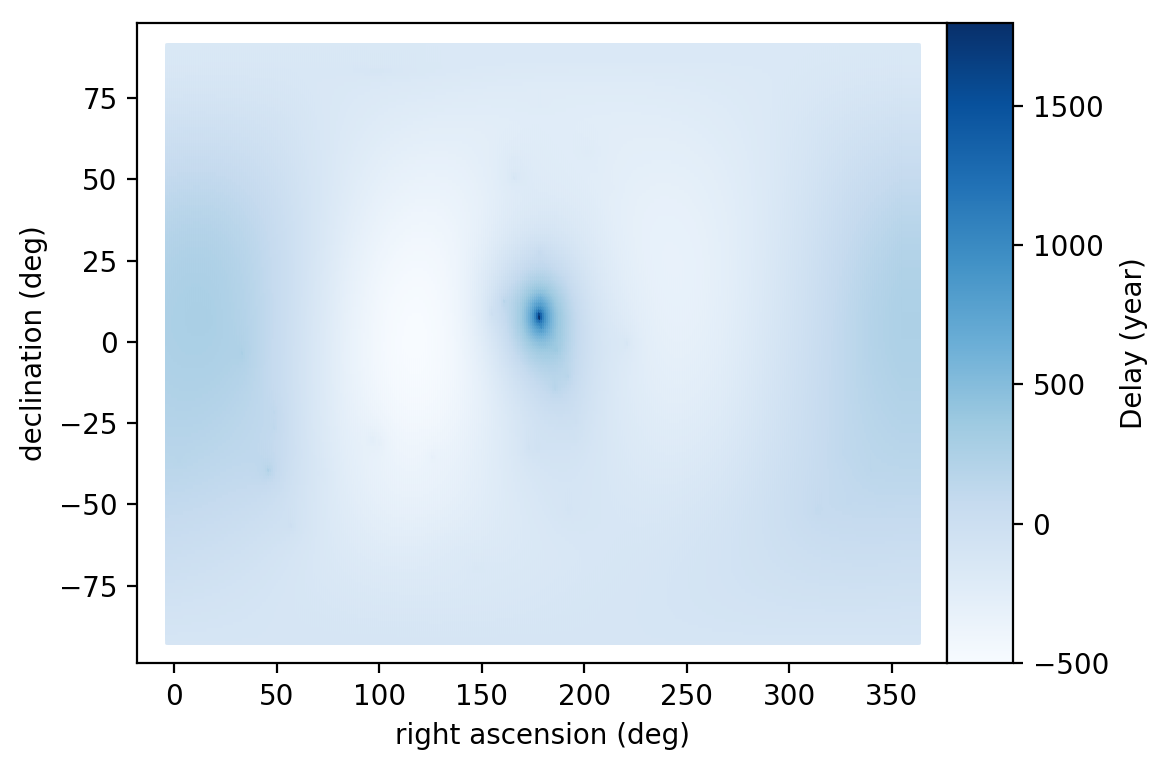}
\end{subfigure}
\caption{Sky maps of the Shapiro delay for sources all over the sky at a given distance (26 Mpc) estimated using the different catalogs we consider.}
\label{img:Shapiro_sky}
\end{figure}

Also, one can see in Fig.~\ref{img:Shapiro_sky} that the Shapiro delay can be close, or even equal, to zero for several locations of the source of the gravitational and electromagnetic waves.

Although the results from the two catalogs are different in magnitude, one can see in Figs.~\ref{img:Shapiro_wrt_size} and \ref{img:Shapiro_sky} that their behaviors are roughly consistent. However, from the different mass estimate used in Tully2015, one can see that there are also significant variations that are caused by different mass estimate, in addition to different catalog completeness.

\section{Conclusion}
\label{sec:conc}

The standard constraints on the equivalence principle using the arrival times of astronomical messengers with different properties emitted in close succession from the same source  suffer from a variety of issues. The most basic of these is that the Shapiro delay considered is not an observable in general relativity, and it is not clear how to relate the standard gauge-dependent way it is calculated to equivalence principle violating alternative theories. However, even assuming that the standard calculation is a useful way of constraining some class of alternative theories, there are further problems encountered when one applies these calculations on cosmological scales, as is commonly done. In this paper, we describe these issues. 

We first use a toy model in order to show arithmetically the appearance of a divergence in the metric perturbation used to compute the Shapiro delay when one is using the set of gauge restrictions [Eqs.~(\ref{eq:usual_gauge}-\ref{eq:usual_gauge2})] that is commonly implicitly used in the literature. We show that the divergence appears because of an unsuitable choice of coordinate time in order to describe the metric, and is therefore simply cured by using an appropriate (non-ambiguous) coordinate time---for instance, a coordinate time that corresponds to the proper time of any given observer.

However, the resulting expression for the Shapiro delay after fixing the coordinate time issue is no longer monotonic with the number of the gravitational sources. This thus prevents one from estimating a conservative minimum amplitude of the Shapiro delay from a subset of the sources of the gravitational field. As a consequence, analyses based on Eq.~(\ref{eq:gen_usual_shapiro}) are not conservative on cosmological scales.\footnote{However, from Eq.~(\ref{eq:Shapiro_kepl}), one can verify that the issue, of course, does not appear for Solar System calculations because $r_P \gg R_{EO}$ for all the significant gravitational sources that are located outside the Solar System. In other words, what makes the two cases (cosmological versus Solar System) different is the distance traveled by light.}

While it might be possible to constrain the gravitational field along the line of sight using cosmological observations (e.g., of galaxy velocities, as in~\cite{Hoffman:2017ako}), and thus avoid the need to use Eq.~(\ref{eq:gen_usual_shapiro}), we do not consider this here. We feel that any further developments of this test should be informed by a fundamental theory, to avoid the gauge dependence of the current formulation of the test, and hope that this paper encourages such work.

\begin{acknowledgments}
The authors thank Shantanu Desai, Daniel Holz, and Robert Wald for useful comments while preparing the final version of the manuscript and Francesco Pannarale for a careful reading of the manuscript. O.M. thanks Aur\'elien Hees for his insightful feedback on a preliminary version of the manuscript.
This research has made use of the VizieR catalogue access tool, CDS,
Strasbourg, France~\cite{Vizier}. The original description
of the VizieR service was published in~\cite{Ochsenbein:2000th}.
N.~K.~J.-M.\ acknowledges support from STFC Consolidator Grant No.~ST/L000636/1. Also, this work has received funding from the European Union's Horizon 2020 research and innovation programme under the Marie Sk{\l}odowska-Curie Grant Agreement No.~690904. 
M.S. is supported in part by STFC grant ST/P000258/1.
This document is LIGO-P1900149-v3.
\end{acknowledgments}

\appendix

\section{Computing the affine distance in the post-Newtonian metric}
\label{app:affine}

The affine distance (see, e.g., Sec.~2.4 of~\cite{Perlick:2004tq}) is defined for a given null geodesic and observer by the affine parameter $\lambda$ of the past-oriented geodesic in the specific parameterization where $\lambda = 0$ corresponds to the spacetime event of the observation and $g_{\mu\nu}\dot{x}^\mu(0) u_0^\nu = 1$, where Greek letters denote spacetime indices, $x^\mu(\lambda)$ is the geodesic, $u_0^\nu$ is the observer's 4-velocity at the observation, and overdots denote derivatives with respect to the affine parameter.

To calculate this distance for the post-Newtonian metric, we start from the post-Newtonian line element given in Eqs.~(7.104) of~\cite{2014grav.book.....P}, which we take through $\mathcal{O}(c^{-3})$ and omit the remainders:
\<\label{eq:metric}
\begin{split}
ds^2 &= g_{\mu\nu}dx^\mu dx^\nu\\
&= -[1 - 2\epsilon U(\vec{x})]dt^2 - 8\epsilon^{3/2}U_k(\vec{x})dtdx^k\\
&\quad + [1 + 2\epsilon U(\vec{x})]\|d\vec{x}\|^2,
\end{split}
\?
where we have set $G = c = 1$ (except for order counting in powers of $c^{-1}$), and introduce $\epsilon = \mathcal{O}(c^{-2})$ as our order counting parameter.
Here $U$ is the Newtonian potential and $U_k$ is the vector potential, both defined explicitly in Box~7.5 of~\cite{2014grav.book.....P}.
We use Roman letters to denote spatial indices and raise and lower indices with the Minkowski metric. Additionally, we have taken the time dependence of all these potentials to be negligible, since we want to consider a situation that is stationary to a good approximation. We take the observation to occur at the origin, with $\lambda = 0$, and the source to be at a spatial location of $\vec{d}_\text{source}$, with $\lambda = \lambda_\text{source}$.

Null geodesics in this metric have to satisfy
\begin{widetext}
\begin{subequations}
\label{eq:geodesic}
\begin{gather}
-[1 - 2\epsilon U(\vec{x}(\lambda))]\dot{t}^2(\lambda) - 8\epsilon^{3/2}U_k(\vec{x}(\lambda))\dot{t}(\lambda) \dot{x}^k(\lambda) + [1 + 2\epsilon U(\vec{x}(\lambda))]\|\dot{\vec{x}}(\lambda)\|^2 = 0,\\
2[1 + 2\epsilon U(\vec{x}(\lambda))]\ddot{x}_k(\lambda) - 2\epsilon\partial_k U(\vec{x}(\lambda))[\dot{t}^2(\lambda) + \|\dot{\vec{x}}(\lambda)\|^2] + 4\epsilon \partial_l U(\vec{x}(\lambda))\dot{x}^l(\lambda)\dot{x}_k(\lambda) \nonumber\\
+ 16\epsilon^{3/2}\partial_{[k}U_{l]}(\vec{x}(\lambda))\dot{t}(\lambda)\dot{x}^l(\lambda) - 8\epsilon^{3/2}U_k(\vec{x}(\lambda))\ddot{t}(\lambda) = 0\label{eq:geodesic2}
\end{gather}
\end{subequations}
(the null geodesic condition and the spatial part of the geodesic equation), where we have written the geodesic as $x^\mu(\lambda) = (t(\lambda),\vec{x}(\lambda))$.
We write the geodesic of the metric \eqref{eq:metric} as
\<
x^\mu(\lambda) = \left(t_0(\lambda) + \epsilon t_1(\lambda) + \epsilon^{3/2} t_{1.5}(\lambda),\vec{x}_0(\lambda) +  \epsilon\vec{x}_1(\lambda) +  \epsilon^{3/2}\vec{x}_{1.5}(\lambda)\right).
\?
\end{widetext}
We then expand Eqs.~\eqref{eq:geodesic} to $\epsilon^{3/2}$, starting with the null geodesic condition, which gives (recalling that the geodesic is past-oriented)
\begin{subequations}
\begin{gather}
\dot{t}_0(\lambda) = -\|\dot{\vec{x}}_0(\lambda)\|, \label{eq:null_Mink}\\
\dot{t}_1(\lambda) = -2U(\vec{x}_0(\lambda))\|\dot{\vec{x}}_0(\lambda)\| - \hat{\vec{n}}_0(\lambda)\cdot\dot{\vec{x}}_1(\lambda),\\
\dot{t}_{1.5}(\lambda) = -4\dot{x}^k_0(\lambda)U_k(\vec{x}_0(\lambda)) - \hat{\vec{n}}_0(\lambda)\cdot\dot{\vec{x}}_{1.5}(\lambda),
\end{gather}
\end{subequations}
where $\hat{\vec{n}}_0 := \dot{\vec{x}}_0/\|\dot{\vec{x}}_0\|$ and we have used the lower-order equations freely in simplifying the higher-order equations. Similarly, Eq.~\eqref{eq:geodesic2} gives
\begin{subequations}
\begin{gather}
\ddot{\vec{x}}_0(\lambda) = 0,\\
\ddot{\vec{x}}_1(\lambda) = 2\|\dot{\vec{x}}_0(\lambda)\|^2\vec{\nabla}U(\vec{x}_0(\lambda)) - 2[\dot{\vec{x}}_0(\lambda)\cdot\vec{\nabla}U(\vec{x}_0(\lambda))]\dot{\vec{x}}_0(\lambda),\label{eq:x1_eq}\\
\ddot{x}^k_{1.5}(\lambda) = 8\partial_{[k}U_{l]}(\vec{x}_0(\lambda))\dot{x}_0^l(\lambda)\|\dot{\vec{x}}_0(\lambda)\|.
\end{gather}
\end{subequations}
From these expressions, we see that $\dot{\vec{x}}_0$ is constant, and $\dot{\vec{x}}_0\cdot\ddot{\vec{x}}_1 = \dot{\vec{x}}_0\cdot\ddot{\vec{x}}_{1.5} = 0$. We take $\vec{x}_0(0) = \vec{0}$ and $\vec{x}_0(\lambda_\text{source}) = \vec{d}_\text{source}$. Thus, we have $\dot{\vec{x}}_0\cdot\dot{\vec{x}}_A = \text{const.}$, for $A\in\{1,1.5\}$, where the constant has to be zero, because the perturbation to the path should not change its endpoints, so $\vec{x}_A(0) = \vec{x}_A(\lambda_\text{source}) = 0$. Thus, $\dot{\vec{x}}_0\cdot\dot{\vec{x}}_1 = \dot{\vec{x}}_0\cdot\dot{\vec{x}}_{1.5} = 0$.

Thus, for $u_0^\mu = (\partial/\partial t)^\mu$ (i.e., an observer at rest with respect to the background Minkowski metric)
\<
g_{\mu\nu}\dot{x}^\mu(0) u_0^\nu = \|\dot{\vec{x}}_0(0)\| + \mathcal{O}(\epsilon^2),
\?
where we have included the remainder term explicitly, to emphasize the order to which this holds.

This means that we want to scale $\lambda$ by $\|\dot{\vec{x}}_0\|$ (recalling that $\dot{\vec{x}}_0$ is constant). Thus, with the scaled $\lambda$, $\dot{\vec{x}}_0 = \vec{d}_\text{source}/\|\vec{d}_\text{source}\|$ and we have $d_\text{aff} = \lambda_\text{source} = \|\vec{d}_\text{source}\|$. The affine distance is thus the distance one would na{\"\i}vely compute using the background metric.

\section{Convergence of lattice sums}
\label{app:convergence}

Here we show that the lattice sums in Eqs.~\eqref{eq:tidal_pot_def} and~\eqref{eq:tidal_pot_grad_sum} converge when summed on expanding cubes or spheres (centered at the origin). We first note that these series are not absolutely convergent, since the magnitudes of their terms do not fall off faster than $\|\vec{\beta}\|^{-3}$ as $\|\vec{\beta}\|\to\infty$. Thus, their convergence (and value) depends on the order in which they are summed (see, e.g.,~\cite{1985JMP....26.2999B} for a discussion of this for the Madelung constant that gives the binding energy of an ion of NaCl). An obvious way to sum is using expanding cubes, since this method retains many of the symmetries of the underlying lattice. However, this is not the only method of computing the lattice sum for which it converges; unlike for the Madelung constant (which is more subtle, due to the alternating nature of the function being summed), a sum using expanding spheres would also converge, though a sum using regions with less symmetry, such as expanding rectangular boxes, would generically not converge.

To demonstrate convergence of the sum in Eq.~\eqref{eq:tidal_pot_def}, we apply Taylor's theorem with Lagrange remainder to
\be
f(\alpha) :=  \frac{1}{\|\alpha\vec \xi - \vec \beta\|} - \frac{1}{ \|\vec \beta\|},
\ee
yielding
\begin{widetext}
\be
\frac{1}{\|\vec \xi - \vec \beta\|} - \frac{1}{ \|\vec \beta\|} = f(1) = \frac{\vec{\xi}\cdot\vec{\beta}}{\|\vec{\beta}\|^3} + \frac{3(\vec{\xi}\cdot\vec{\beta})^2 - \|\vec{\xi}\|^2\|\vec{\beta}\|^2}{\|\vec \beta\|^5}
+ \frac{(\bar{\alpha}\|\vec{\xi}\|^2 - \vec{\xi}\cdot\vec{\beta})[3\|\vec{\xi}\|^2\|\bar{\alpha}\vec{\xi} - \vec{\beta}\|^2 - 5(\vec{\xi}\cdot\vec{\beta} - \bar{\alpha}\|\vec{\xi}\|^2)^2]}{2\|\bar{\alpha}\vec{\xi} - \vec{\beta}\|^7}
\ee
\end{widetext}
for some $\bar{\alpha}\in(0,1)$ (depending on $\vec{\xi}$ and $\vec{\beta}$). The first two terms vanish when summed over a cube (or sphere) centered at the origin. The first term vanishes because the set of points in the cube (sphere) centered at the origin is symmetric under $\vec{\beta} \to -\vec{\beta}$. The second term vanishes because the set of points in the cube (or sphere) centered at the origin is symmetric under $\beta_{1,2,3} \to -\beta_{1,2,3}$ (i.e., switching the sign of the individual components of $\vec{\beta}$) and is also symmetric under cyclic permutations of the indices. The first of these symmetries implies that the $\beta_k\beta_l$ ($k\neq l$) terms vanish upon summation, giving a summand of $[3(\xi_1^2\beta_1^2 + \xi_2^2\beta_2^2 + \xi_3^2\beta_3^2) - \|\vec{\xi}\|^2\|\vec{\beta}\|^2]/\|\vec{\beta}\|^5$. This vanishes when summed over all cyclic permutations of indices.

We now want to bound the remainder term. Since we are only interested in its behavior for large $\|\vec{\beta}\|$, we can assume that $\|\vec{\beta}\|\geq 2\|\vec{\xi}\|$. We then apply the Cauchy-Schwarz inequality and triangle inequality (a number of times) to the numerator to bound its magnitude by $(\|\vec{\xi}\|^2 + \|\vec{\xi}\|\|\vec{\beta}\|)[3\|\vec{\xi}\|^2(\|\vec{\xi}\| + \|\vec{\beta}\|)^2 + 5(\|\vec{\xi}\|\|\vec{\beta}\| + \|\vec{\xi}\|^2)^2] \leq 27\|\vec{\xi}\|^3\|\vec{\beta}\|^3$. We use the assumption that $\|\vec{\xi}\|\leq\|\vec{\beta}\|/2$ to obtain the final bound. We also apply the reverse triangle inequality to give a lower bound on the magnitude of the denominator of $\|\vec{\beta}\|^7/64$, and thus an upper bound of $1728\|\vec{\xi}\|^3/\|\vec{\beta}\|^4$ on the magnitude of the remainder.\footnote{This is not a sharp bound, but suffices for our purposes.} The sum of this bound over $\vec{\beta}\in \mathbb{Z}^3_*$ converges, thus demonstrating that the sum on expanding cubes or spheres of Eq.~\eqref{eq:tidal_pot_def} indeed converges.

We now apply the same technique to demonstrate that the sum in Eq.~\eqref{eq:tidal_pot_grad_sum} converges, i.e., we note that Taylor's theorem with Lagrange remainder gives
\begin{widetext}
\be
\frac{\xi^i - \beta^i}{\|\vec \xi - \vec \beta\|^3} = -\frac{\beta^i}{\|\vec \beta\|^3} + \frac{\|\vec{\beta}\|^2\xi^i - 3(\vec{\xi}\cdot\vec{\beta})\beta^i}{\|\vec \beta\|^5}
+ \frac{2\|\bar{\alpha}\vec{\xi} - \vec{\beta}\|^2(\vec{\xi}\cdot\vec{\beta} - \bar{\alpha}\|\vec{\xi}\|^2)\xi^i - [\|\vec{\xi}\|^2\|\bar{\alpha}\vec{\xi} - \vec{\beta}\|^2 - 5(\vec{\xi}\cdot\vec{\beta} - \bar{\alpha}\|\vec{\xi}\|^2)^2](\bar{\alpha}\xi^i - \beta^i)}{2\|\bar{\alpha}\vec{\xi} - \vec{\beta}\|^7}
\ee
\end{widetext}
for some $\bar{\alpha}\in(0,1)$ (depending on $\vec{\xi}$ and $\vec{\beta}$). As before, we find that the first two terms vanish when summed over a cube or sphere centered at the origin. This is obvious for the first term. For the second term, we first note that the sums of the $\beta_k\beta_l$ ($k\neq l$) terms (over a cube or sphere centered at the origin) vanish, giving $[\|\vec{\beta}\|^2 - 3(\beta^i)^2]\xi^i/\|\vec \beta\|^5$ (no sum), which vanishes when summed over all cyclic permutations of indices. We can bound the remainder using the same techniques as before, along with noting that $|\zeta^i| \leq \|\vec{\zeta}\|$ for any vector $\vec{\zeta}$, giving an upper bound for the numerator of $2(\|\vec{\xi}\| + \|\vec{\beta}\|)^2(\|\vec{\xi}\|\|\vec{\beta}\| + \|\vec{\xi}\|^2)\|\vec{\xi}\| + [\|\vec{\xi}\|^2(\|\vec{\xi}\| + \|\vec{\beta}\|)^2 + 5(\|\vec{\xi}\|\|\vec{\beta}\| + \|\vec{\xi}\|^2)^2](\|\vec{\xi}\| + \|\vec{\beta}\|) \leq 27\|\vec{\xi}\|^2\|\vec{\beta}\|^3$, for $\|\vec{\xi}\|\leq\|\vec{\beta}\|/2$, as before. The lower bound on the denominator is the same as before, giving an overall upper bound on the magnitude of the remainder of $1728\|\vec{\xi}\|^2/\|\vec{\beta}\|^4$, which converges when summed over $\vec{\beta}\in \mathbb{Z}^3_*$, thus demonstrating that the sum on expanding cubes or spheres of Eq.~\eqref{eq:tidal_pot_grad_sum} indeed converges.


\begin{thebibliography}{84}%
\makeatletter
\providecommand \@ifxundefined [1]{%
 \@ifx{#1\undefined}
}%
\providecommand \@ifnum [1]{%
 \ifnum #1\expandafter \@firstoftwo
 \else \expandafter \@secondoftwo
 \fi
}%
\providecommand \@ifx [1]{%
 \ifx #1\expandafter \@firstoftwo
 \else \expandafter \@secondoftwo
 \fi
}%
\providecommand \natexlab [1]{#1}%
\providecommand \enquote  [1]{``#1''}%
\providecommand \bibnamefont  [1]{#1}%
\providecommand \bibfnamefont [1]{#1}%
\providecommand \citenamefont [1]{#1}%
\providecommand \href@noop [0]{\@secondoftwo}%
\providecommand \href [0]{\begingroup \@sanitize@url \@href}%
\providecommand \@href[1]{\@@startlink{#1}\@@href}%
\providecommand \@@href[1]{\endgroup#1\@@endlink}%
\providecommand \@sanitize@url [0]{\catcode `\\12\catcode `\$12\catcode
  `\&12\catcode `\#12\catcode `\^12\catcode `\_12\catcode `\%12\relax}%
\providecommand \@@startlink[1]{}%
\providecommand \@@endlink[0]{}%
\providecommand \url  [0]{\begingroup\@sanitize@url \@url }%
\providecommand \@url [1]{\endgroup\@href {#1}{\urlprefix }}%
\providecommand \urlprefix  [0]{URL }%
\providecommand \Eprint [0]{\href }%
\providecommand \doibase [0]{http://dx.doi.org/}%
\providecommand \selectlanguage [0]{\@gobble}%
\providecommand \bibinfo  [0]{\@secondoftwo}%
\providecommand \bibfield  [0]{\@secondoftwo}%
\providecommand \translation [1]{[#1]}%
\providecommand \BibitemOpen [0]{}%
\providecommand \bibitemStop [0]{}%
\providecommand \bibitemNoStop [0]{.\EOS\space}%
\providecommand \EOS [0]{\spacefactor3000\relax}%
\providecommand \BibitemShut  [1]{\csname bibitem#1\endcsname}%
\let\auto@bib@innerbib\@empty
\bibitem [{\citenamefont {Isaacson}(1968)}]{Isaacson:1967zz}%
  \BibitemOpen
  \bibfield  {author} {\bibinfo {author} {\bibfnamefont {R.~A.}\ \bibnamefont
  {Isaacson}},\ }\bibfield  {title} {\enquote {\bibinfo {title} {{Gravitational
  Radiation in the Limit of High Frequency. I. The Linear Approximation and
  Geometrical Optics}},}\ }\href {\doibase 10.1103/PhysRev.166.1263} {\bibfield
   {journal} {\bibinfo  {journal} {Phys. Rev.}\ }\textbf {\bibinfo {volume}
  {166}},\ \bibinfo {pages} {1263} (\bibinfo {year} {1968})}\BibitemShut
  {NoStop}%
\bibitem [{\citenamefont {{Takahashi}}(2017)}]{takahashi:2017aj}%
  \BibitemOpen
  \bibfield  {author} {\bibinfo {author} {\bibfnamefont {R.}~\bibnamefont
  {{Takahashi}}},\ }\bibfield  {title} {\enquote {\bibinfo {title} {{Arrival
  Time Differences between Gravitational Waves and Electromagnetic Signals due
  to Gravitational Lensing}},}\ }\href {\doibase 10.3847/1538-4357/835/1/103}
  {\bibfield  {journal} {\bibinfo  {journal} {Astrophys. J.}\ }\textbf
  {\bibinfo {volume} {835}},\ \bibinfo {eid} {103} (\bibinfo {year} {2017})},\
  \Eprint {http://arxiv.org/abs/1606.00458} {arXiv:1606.00458 [astro-ph.CO]}
  \BibitemShut {NoStop}%
\bibitem [{\citenamefont {{Sivaram}}(1999)}]{1999BASI...27..627S}%
  \BibitemOpen
  \bibfield  {author} {\bibinfo {author} {\bibfnamefont {C.}~\bibnamefont
  {{Sivaram}}},\ }\bibfield  {title} {\enquote {\bibinfo {title} {{Constraints
  on the photon mass and charge and test of equivalence principle from GRB
  990123}},}\ }\href@noop {} {\bibfield  {journal} {\bibinfo  {journal} {Bull.
  Astron. Soc. India}\ }\textbf {\bibinfo {volume} {27}},\ \bibinfo {pages}
  {627} (\bibinfo {year} {1999})}\BibitemShut {NoStop}%
\bibitem [{\citenamefont {Desai}\ \emph {et~al.}(2008)\citenamefont {Desai},
  \citenamefont {Kahya},\ and\ \citenamefont {Woodard}}]{Desai:2008vj}%
  \BibitemOpen
  \bibfield  {author} {\bibinfo {author} {\bibfnamefont {S.}~\bibnamefont
  {Desai}}, \bibinfo {author} {\bibfnamefont {E.~O.}\ \bibnamefont {Kahya}}, \
  and\ \bibinfo {author} {\bibfnamefont {R.~P.}\ \bibnamefont {Woodard}},\
  }\bibfield  {title} {\enquote {\bibinfo {title} {{Reduced time delay for
  gravitational waves with dark matter emulators}},}\ }\href {\doibase
  10.1103/PhysRevD.77.124041} {\bibfield  {journal} {\bibinfo  {journal} {Phys.
  Rev. D}\ }\textbf {\bibinfo {volume} {77}},\ \bibinfo {pages} {124041}
  (\bibinfo {year} {2008})},\ \Eprint {http://arxiv.org/abs/0804.3804}
  {arXiv:0804.3804 [astro-ph]} \BibitemShut {NoStop}%
\bibitem [{\citenamefont {{Kahya}}(2011)}]{kahya:2011pl}%
  \BibitemOpen
  \bibfield  {author} {\bibinfo {author} {\bibfnamefont {E.~O.}\ \bibnamefont
  {{Kahya}}},\ }\bibfield  {title} {\enquote {\bibinfo {title} {{A useful guide
  for gravitational wave observers to test modified gravity models}},}\ }\href
  {\doibase 10.1016/j.physletb.2011.05.073} {\bibfield  {journal} {\bibinfo
  {journal} {Phys. Lett. B}\ }\textbf {\bibinfo {volume} {701}},\ \bibinfo
  {pages} {291} (\bibinfo {year} {2011})},\ \Eprint
  {http://arxiv.org/abs/1001.0725} {arXiv:1001.0725 [gr-qc]} \BibitemShut
  {NoStop}%
\bibitem [{\citenamefont {{Gao}}\ \emph {et~al.}(2015)\citenamefont {{Gao}},
  \citenamefont {{Wu}},\ and\ \citenamefont {{M{\'e}sz{\'a}ros}}}]{gao:2015aj}%
  \BibitemOpen
  \bibfield  {author} {\bibinfo {author} {\bibfnamefont {H.}~\bibnamefont
  {{Gao}}}, \bibinfo {author} {\bibfnamefont {X.-F.}\ \bibnamefont {{Wu}}}, \
  and\ \bibinfo {author} {\bibfnamefont {P.}~\bibnamefont
  {{M{\'e}sz{\'a}ros}}},\ }\bibfield  {title} {\enquote {\bibinfo {title}
  {{Cosmic Transients Test Einstein's Equivalence Principle out to GeV
  Energies}},}\ }\href {\doibase 10.1088/0004-637X/810/2/121} {\bibfield
  {journal} {\bibinfo  {journal} {Astrophys. J.}\ }\textbf {\bibinfo {volume}
  {810}},\ \bibinfo {eid} {121} (\bibinfo {year} {2015})},\ \Eprint
  {http://arxiv.org/abs/1509.00150} {arXiv:1509.00150 [astro-ph.HE]}
  \BibitemShut {NoStop}%
\bibitem [{\citenamefont {{Desai}}\ and\ \citenamefont
  {{Kahya}}(2016)}]{desai:2016mp}%
  \BibitemOpen
  \bibfield  {author} {\bibinfo {author} {\bibfnamefont {S.}~\bibnamefont
  {{Desai}}}\ and\ \bibinfo {author} {\bibfnamefont {E.~O.}\ \bibnamefont
  {{Kahya}}},\ }\bibfield  {title} {\enquote {\bibinfo {title} {{Galactic
  one-way Shapiro delay to PSR B1937+21}},}\ }\href {\doibase
  10.1142/S0217732316500838} {\bibfield  {journal} {\bibinfo  {journal} {Mod.
  Phys. Lett. A}\ }\textbf {\bibinfo {volume} {31}},\ \bibinfo {eid} {1650083}
  (\bibinfo {year} {2016})},\ \Eprint {http://arxiv.org/abs/1510.08228}
  {arXiv:1510.08228} \BibitemShut {NoStop}%
\bibitem [{\citenamefont {{Wei}}\ \emph {et~al.}(2015)\citenamefont {{Wei}},
  \citenamefont {{Gao}}, \citenamefont {{Wu}},\ and\ \citenamefont
  {{M{\'e}sz{\'a}ros}}}]{wei:2015pl}%
  \BibitemOpen
  \bibfield  {author} {\bibinfo {author} {\bibfnamefont {J.-J.}\ \bibnamefont
  {{Wei}}}, \bibinfo {author} {\bibfnamefont {H.}~\bibnamefont {{Gao}}},
  \bibinfo {author} {\bibfnamefont {X.-F.}\ \bibnamefont {{Wu}}}, \ and\
  \bibinfo {author} {\bibfnamefont {P.}~\bibnamefont {{M{\'e}sz{\'a}ros}}},\
  }\bibfield  {title} {\enquote {\bibinfo {title} {{Testing Einstein's
  Equivalence Principle With Fast Radio Bursts}},}\ }\href {\doibase
  10.1103/PhysRevLett.115.261101} {\bibfield  {journal} {\bibinfo  {journal}
  {Phys. Rev. Lett.}\ }\textbf {\bibinfo {volume} {115}},\ \bibinfo {eid}
  {261101} (\bibinfo {year} {2015})},\ \Eprint
  {http://arxiv.org/abs/1512.07670} {arXiv:1512.07670 [astro-ph.HE]}
  \BibitemShut {NoStop}%
\bibitem [{\citenamefont {{Li}}\ \emph {et~al.}(2016)\citenamefont {{Li}},
  \citenamefont {{Hu}}, \citenamefont {{Fan}},\ and\ \citenamefont
  {{Wei}}}]{li:2016ap}%
  \BibitemOpen
  \bibfield  {author} {\bibinfo {author} {\bibfnamefont {X.}~\bibnamefont
  {{Li}}}, \bibinfo {author} {\bibfnamefont {Y.-M.}\ \bibnamefont {{Hu}}},
  \bibinfo {author} {\bibfnamefont {Y.-Z.}\ \bibnamefont {{Fan}}}, \ and\
  \bibinfo {author} {\bibfnamefont {D.-M.}\ \bibnamefont {{Wei}}},\ }\bibfield
  {title} {\enquote {\bibinfo {title} {{GRB/GW Association: Long-Short GRB
  Candidates, Time Lag, Measuring Gravitational Wave Velocity, and Testing
  Einstein's Equivalence Principle}},}\ }\href {\doibase
  10.3847/0004-637X/827/1/75} {\bibfield  {journal} {\bibinfo  {journal}
  {Astrophys. J.}\ }\textbf {\bibinfo {volume} {827}},\ \bibinfo {eid} {75}
  (\bibinfo {year} {2016})},\ \Eprint {http://arxiv.org/abs/1601.00180}
  {arXiv:1601.00180 [astro-ph.HE]} \BibitemShut {NoStop}%
\bibitem [{\citenamefont {{Nusser}}(2016)}]{nusser:2016ap}%
  \BibitemOpen
  \bibfield  {author} {\bibinfo {author} {\bibfnamefont {A.}~\bibnamefont
  {{Nusser}}},\ }\bibfield  {title} {\enquote {\bibinfo {title} {{On Testing
  the Equivalence Principle with Extragalactic Bursts}},}\ }\href {\doibase
  10.3847/2041-8205/821/1/L2} {\bibfield  {journal} {\bibinfo  {journal}
  {Astrophys. J. Lett.}\ }\textbf {\bibinfo {volume} {821}},\ \bibinfo {eid}
  {L2} (\bibinfo {year} {2016})},\ \Eprint {http://arxiv.org/abs/1601.03636}
  {arXiv:1601.03636} \BibitemShut {NoStop}%
\bibitem [{\citenamefont {{Wei}}\ \emph
  {et~al.}(2016{\natexlab{a}})\citenamefont {{Wei}}, \citenamefont {{Wang}},
  \citenamefont {{Gao}},\ and\ \citenamefont {{Wu}}}]{wei:2016ap}%
  \BibitemOpen
  \bibfield  {author} {\bibinfo {author} {\bibfnamefont {J.-J.}\ \bibnamefont
  {{Wei}}}, \bibinfo {author} {\bibfnamefont {J.-S.}\ \bibnamefont {{Wang}}},
  \bibinfo {author} {\bibfnamefont {H.}~\bibnamefont {{Gao}}}, \ and\ \bibinfo
  {author} {\bibfnamefont {X.-F.}\ \bibnamefont {{Wu}}},\ }\bibfield  {title}
  {\enquote {\bibinfo {title} {{Tests of the Einstein Equivalence Principle
  Using TeV Blazars}},}\ }\href {\doibase 10.3847/2041-8205/818/1/L2}
  {\bibfield  {journal} {\bibinfo  {journal} {Astrophys. J. Lett.}\ }\textbf
  {\bibinfo {volume} {818}},\ \bibinfo {eid} {L2} (\bibinfo {year}
  {2016}{\natexlab{a}})},\ \Eprint {http://arxiv.org/abs/1601.04145}
  {arXiv:1601.04145 [astro-ph.HE]} \BibitemShut {NoStop}%
\bibitem [{\citenamefont {{Wu}}\ \emph {et~al.}(2016)\citenamefont {{Wu}},
  \citenamefont {{Gao}}, \citenamefont {{Wei}}, \citenamefont
  {{M{\'e}sz{\'a}ros}}, \citenamefont {{Zhang}}, \citenamefont {{Dai}},
  \citenamefont {{Zhang}},\ and\ \citenamefont {{Zhu}}}]{wu:2016pr}%
  \BibitemOpen
  \bibfield  {author} {\bibinfo {author} {\bibfnamefont {X.-F.}\ \bibnamefont
  {{Wu}}}, \bibinfo {author} {\bibfnamefont {H.}~\bibnamefont {{Gao}}},
  \bibinfo {author} {\bibfnamefont {J.-J.}\ \bibnamefont {{Wei}}}, \bibinfo
  {author} {\bibfnamefont {P.}~\bibnamefont {{M{\'e}sz{\'a}ros}}}, \bibinfo
  {author} {\bibfnamefont {B.}~\bibnamefont {{Zhang}}}, \bibinfo {author}
  {\bibfnamefont {Z.-G.}\ \bibnamefont {{Dai}}}, \bibinfo {author}
  {\bibfnamefont {S.-N.}\ \bibnamefont {{Zhang}}}, \ and\ \bibinfo {author}
  {\bibfnamefont {Z.-H.}\ \bibnamefont {{Zhu}}},\ }\bibfield  {title} {\enquote
  {\bibinfo {title} {{Testing Einstein's weak equivalence principle with
  gravitational waves}},}\ }\href {\doibase 10.1103/PhysRevD.94.024061}
  {\bibfield  {journal} {\bibinfo  {journal} {\prd}\ }\textbf {\bibinfo
  {volume} {94}},\ \bibinfo {eid} {024061} (\bibinfo {year} {2016})},\ \Eprint
  {http://arxiv.org/abs/1602.01566} {arXiv:1602.01566 [astro-ph.HE]}
  \BibitemShut {NoStop}%
\bibitem [{\citenamefont {{Kahya}}\ and\ \citenamefont
  {{Desai}}(2016)}]{kahya:2016ph}%
  \BibitemOpen
  \bibfield  {author} {\bibinfo {author} {\bibfnamefont {E.~O.}\ \bibnamefont
  {{Kahya}}}\ and\ \bibinfo {author} {\bibfnamefont {S.}~\bibnamefont
  {{Desai}}},\ }\bibfield  {title} {\enquote {\bibinfo {title} {{Constraints on
  frequency-dependent violations of Shapiro delay from GW150914}},}\ }\href
  {\doibase 10.1016/j.physletb.2016.03.033} {\bibfield  {journal} {\bibinfo
  {journal} {Phys. Lett. B}\ }\textbf {\bibinfo {volume} {756}},\ \bibinfo
  {pages} {265} (\bibinfo {year} {2016})},\ \Eprint
  {http://arxiv.org/abs/1602.04779} {arXiv:1602.04779 [gr-qc]} \BibitemShut
  {NoStop}%
\bibitem [{\citenamefont {{Wang}}\ \emph {et~al.}(2016)\citenamefont {{Wang}},
  \citenamefont {{Liu}},\ and\ \citenamefont {{Wang}}}]{wang:2016pl}%
  \BibitemOpen
  \bibfield  {author} {\bibinfo {author} {\bibfnamefont {Z.-Y.}\ \bibnamefont
  {{Wang}}}, \bibinfo {author} {\bibfnamefont {R.-Y.}\ \bibnamefont {{Liu}}}, \
  and\ \bibinfo {author} {\bibfnamefont {X.-Y.}\ \bibnamefont {{Wang}}},\
  }\bibfield  {title} {\enquote {\bibinfo {title} {{Testing the Equivalence
  Principle and Lorentz Invariance with PeV Neutrinos from Blazar Flares}},}\
  }\href {\doibase 10.1103/PhysRevLett.116.151101} {\bibfield  {journal}
  {\bibinfo  {journal} {Phys. Rev. Lett.}\ }\textbf {\bibinfo {volume} {116}},\
  \bibinfo {eid} {151101} (\bibinfo {year} {2016})},\ \Eprint
  {http://arxiv.org/abs/1602.06805} {arXiv:1602.06805 [astro-ph.HE]}
  \BibitemShut {NoStop}%
\bibitem [{\citenamefont {{Tingay}}\ and\ \citenamefont
  {{Kaplan}}(2016)}]{tingay:2016ap}%
  \BibitemOpen
  \bibfield  {author} {\bibinfo {author} {\bibfnamefont {S.~J.}\ \bibnamefont
  {{Tingay}}}\ and\ \bibinfo {author} {\bibfnamefont {D.~L.}\ \bibnamefont
  {{Kaplan}}},\ }\bibfield  {title} {\enquote {\bibinfo {title} {{Limits on
  Einstein's Equivalence Principle from the First Localized Fast Radio Burst
  FRB 150418}},}\ }\href {\doibase 10.3847/2041-8205/820/2/L31} {\bibfield
  {journal} {\bibinfo  {journal} {Astrophys. J. Lett.}\ }\textbf {\bibinfo
  {volume} {820}},\ \bibinfo {eid} {L31} (\bibinfo {year} {2016})},\ \Eprint
  {http://arxiv.org/abs/1602.07643} {arXiv:1602.07643} \BibitemShut {NoStop}%
\bibitem [{\citenamefont {{Wei}}\ \emph
  {et~al.}(2016{\natexlab{b}})\citenamefont {{Wei}}, \citenamefont {{Wu}},
  \citenamefont {{Gao}},\ and\ \citenamefont
  {{M{\'e}sz{\'a}ros}}}]{wei:2016jc}%
  \BibitemOpen
  \bibfield  {author} {\bibinfo {author} {\bibfnamefont {J.-J.}\ \bibnamefont
  {{Wei}}}, \bibinfo {author} {\bibfnamefont {X.-F.}\ \bibnamefont {{Wu}}},
  \bibinfo {author} {\bibfnamefont {H.}~\bibnamefont {{Gao}}}, \ and\ \bibinfo
  {author} {\bibfnamefont {P.}~\bibnamefont {{M{\'e}sz{\'a}ros}}},\ }\bibfield
  {title} {\enquote {\bibinfo {title} {{Limits on the neutrino velocity,
  Lorentz invariance, and the weak equivalence principle with TeV neutrinos
  from gamma-ray bursts}},}\ }\href {\doibase 10.1088/1475-7516/2016/08/031}
  {\bibfield  {journal} {\bibinfo  {journal} {J. Cosmol. Astropart. Phys.}\
  }\textbf {\bibinfo {volume} {8}},\ \bibinfo {eid} {031} (\bibinfo {year}
  {2016}{\natexlab{b}})},\ \Eprint {http://arxiv.org/abs/1603.07568}
  {arXiv:1603.07568 [astro-ph.HE]} \BibitemShut {NoStop}%
\bibitem [{\citenamefont {{Liu}}\ \emph {et~al.}(2017)\citenamefont {{Liu}},
  \citenamefont {{Zhao}}, \citenamefont {{You}}, \citenamefont {{Lu}},\ and\
  \citenamefont {{Xu}}}]{liu:2017pl}%
  \BibitemOpen
  \bibfield  {author} {\bibinfo {author} {\bibfnamefont {M.}~\bibnamefont
  {{Liu}}}, \bibinfo {author} {\bibfnamefont {Z.}~\bibnamefont {{Zhao}}},
  \bibinfo {author} {\bibfnamefont {X.}~\bibnamefont {{You}}}, \bibinfo
  {author} {\bibfnamefont {J.}~\bibnamefont {{Lu}}}, \ and\ \bibinfo {author}
  {\bibfnamefont {L.}~\bibnamefont {{Xu}}},\ }\bibfield  {title} {\enquote
  {\bibinfo {title} {{Test of the Weak Equivalence Principle using LIGO
  observations of GW150914 and Fermi observations of GBM transient 150914}},}\
  }\href {\doibase 10.1016/j.physletb.2017.04.033} {\bibfield  {journal}
  {\bibinfo  {journal} {Phys. Lett. B}\ }\textbf {\bibinfo {volume} {770}},\
  \bibinfo {pages} {8} (\bibinfo {year} {2017})},\ \Eprint
  {http://arxiv.org/abs/1604.06668} {arXiv:1604.06668 [gr-qc]} \BibitemShut
  {NoStop}%
\bibitem [{\citenamefont {{Luo}}\ \emph {et~al.}(2016)\citenamefont {{Luo}},
  \citenamefont {{Zhang}}, \citenamefont {{Wei}},\ and\ \citenamefont
  {{Wu}}}]{luo:2016jh}%
  \BibitemOpen
  \bibfield  {author} {\bibinfo {author} {\bibfnamefont {Z.-X.}\ \bibnamefont
  {{Luo}}}, \bibinfo {author} {\bibfnamefont {B.}~\bibnamefont {{Zhang}}},
  \bibinfo {author} {\bibfnamefont {J.-J.}\ \bibnamefont {{Wei}}}, \ and\
  \bibinfo {author} {\bibfnamefont {X.-F.}\ \bibnamefont {{Wu}}},\ }\bibfield
  {title} {\enquote {\bibinfo {title} {{Testing Einstein's Equivalence
  Principle with supercluster Laniakea's gravitational field}},}\ }\href
  {\doibase 10.1016/j.jheap.2016.04.001} {\bibfield  {journal} {\bibinfo
  {journal} {J. High Energy Astrophys.}\ }\textbf {\bibinfo {volume} {9}},\
  \bibinfo {pages} {35} (\bibinfo {year} {2016})},\ \Eprint
  {http://arxiv.org/abs/1604.02566} {arXiv:1604.02566 [astro-ph.HE]}
  \BibitemShut {NoStop}%
\bibitem [{\citenamefont {{Sang}}\ \emph {et~al.}(2016)\citenamefont {{Sang}},
  \citenamefont {{Lin}},\ and\ \citenamefont {{Chang}}}]{sang:2016mn}%
  \BibitemOpen
  \bibfield  {author} {\bibinfo {author} {\bibfnamefont {Y.}~\bibnamefont
  {{Sang}}}, \bibinfo {author} {\bibfnamefont {H.-N.}\ \bibnamefont {{Lin}}}, \
  and\ \bibinfo {author} {\bibfnamefont {Z.}~\bibnamefont {{Chang}}},\
  }\bibfield  {title} {\enquote {\bibinfo {title} {{Testing Einstein's
  equivalence principle with short gamma-ray bursts}},}\ }\href {\doibase
  10.1093/mnras/stw1136} {\bibfield  {journal} {\bibinfo  {journal} {Mon. Not.
  R. Astron. Soc.}\ }\textbf {\bibinfo {volume} {460}},\ \bibinfo {pages}
  {2282} (\bibinfo {year} {2016})},\ \Eprint {http://arxiv.org/abs/1605.02834}
  {arXiv:1605.02834 [astro-ph.HE]} \BibitemShut {NoStop}%
\bibitem [{\citenamefont {{Yang}}\ and\ \citenamefont
  {{Zhang}}(2016)}]{yang:2016pr}%
  \BibitemOpen
  \bibfield  {author} {\bibinfo {author} {\bibfnamefont {Y.-P.}\ \bibnamefont
  {{Yang}}}\ and\ \bibinfo {author} {\bibfnamefont {B.}~\bibnamefont
  {{Zhang}}},\ }\bibfield  {title} {\enquote {\bibinfo {title} {{Testing
  Einstein's weak equivalence principle with a 0.4-nanosecond giant pulse of
  the Crab pulsar}},}\ }\href {\doibase 10.1103/PhysRevD.94.101501} {\bibfield
  {journal} {\bibinfo  {journal} {\prd}\ }\textbf {\bibinfo {volume} {94}},\
  \bibinfo {eid} {101501} (\bibinfo {year} {2016})},\ \Eprint
  {http://arxiv.org/abs/1608.07657} {arXiv:1608.07657 [astro-ph.HE]}
  \BibitemShut {NoStop}%
\bibitem [{\citenamefont {{Zhang}}\ and\ \citenamefont
  {{Gong}}(2017)}]{zhang:2017ap}%
  \BibitemOpen
  \bibfield  {author} {\bibinfo {author} {\bibfnamefont {Y.}~\bibnamefont
  {{Zhang}}}\ and\ \bibinfo {author} {\bibfnamefont {B.}~\bibnamefont
  {{Gong}}},\ }\bibfield  {title} {\enquote {\bibinfo {title} {{Test of Weak
  Equivalence Principle with the Multi-band Timing of the Crab Pulsar}},}\
  }\href {\doibase 10.3847/1538-4357/aa61fb} {\bibfield  {journal} {\bibinfo
  {journal} {Astrophys. J.}\ }\textbf {\bibinfo {volume} {837}},\ \bibinfo
  {eid} {134} (\bibinfo {year} {2017})},\ \Eprint
  {http://arxiv.org/abs/1612.00717} {arXiv:1612.00717 [gr-qc]} \BibitemShut
  {NoStop}%
\bibitem [{\citenamefont {{Desai}}\ and\ \citenamefont
  {{Kahya}}(2018)}]{desai:2018ep}%
  \BibitemOpen
  \bibfield  {author} {\bibinfo {author} {\bibfnamefont {S.}~\bibnamefont
  {{Desai}}}\ and\ \bibinfo {author} {\bibfnamefont {E.}~\bibnamefont
  {{Kahya}}},\ }\bibfield  {title} {\enquote {\bibinfo {title} {{Galactic
  Shapiro delay to the Crab pulsar and limit on weak equivalence principle
  violation}},}\ }\href {\doibase 10.1140/epjc/s10052-018-5571-0} {\bibfield
  {journal} {\bibinfo  {journal} {Eur. Phys. J. C}\ }\textbf {\bibinfo {volume}
  {78}},\ \bibinfo {eid} {86} (\bibinfo {year} {2018})},\ \Eprint
  {http://arxiv.org/abs/1612.02532} {arXiv:1612.02532 [astro-ph.HE]}
  \BibitemShut {NoStop}%
\bibitem [{\citenamefont {Wu}\ \emph {et~al.}(2017)\citenamefont {Wu},
  \citenamefont {Wei}, \citenamefont {Lan}, \citenamefont {Gao}, \citenamefont
  {Dai},\ and\ \citenamefont {M{\'e}sz{\'a}ros}}]{Wu:2017yjl}%
  \BibitemOpen
  \bibfield  {author} {\bibinfo {author} {\bibfnamefont {X.-F.}\ \bibnamefont
  {Wu}}, \bibinfo {author} {\bibfnamefont {J.-J.}\ \bibnamefont {Wei}},
  \bibinfo {author} {\bibfnamefont {M.-X.}\ \bibnamefont {Lan}}, \bibinfo
  {author} {\bibfnamefont {H.}~\bibnamefont {Gao}}, \bibinfo {author}
  {\bibfnamefont {Z.-G.}\ \bibnamefont {Dai}}, \ and\ \bibinfo {author}
  {\bibfnamefont {P.}~\bibnamefont {M{\'e}sz{\'a}ros}},\ }\bibfield  {title}
  {\enquote {\bibinfo {title} {{New test of weak equivalence principle using
  polarized light from astrophysical events}},}\ }\href {\doibase
  10.1103/PhysRevD.95.103004} {\bibfield  {journal} {\bibinfo  {journal} {Phys.
  Rev. D}\ }\textbf {\bibinfo {volume} {95}},\ \bibinfo {pages} {103004}
  (\bibinfo {year} {2017})},\ \Eprint {http://arxiv.org/abs/1703.09935}
  {arXiv:1703.09935 [astro-ph.HE]} \BibitemShut {NoStop}%
\bibitem [{\citenamefont {{Yang}}\ \emph {et~al.}(2017)\citenamefont {{Yang}},
  \citenamefont {{Zou}}, \citenamefont {{Zhang}}, \citenamefont {{Liao}},\ and\
  \citenamefont {{Lei}}}]{yang:2017mn}%
  \BibitemOpen
  \bibfield  {author} {\bibinfo {author} {\bibfnamefont {C.}~\bibnamefont
  {{Yang}}}, \bibinfo {author} {\bibfnamefont {Y.-C.}\ \bibnamefont {{Zou}}},
  \bibinfo {author} {\bibfnamefont {Y.-Y.}\ \bibnamefont {{Zhang}}}, \bibinfo
  {author} {\bibfnamefont {B.}~\bibnamefont {{Liao}}}, \ and\ \bibinfo {author}
  {\bibfnamefont {W.-H.}\ \bibnamefont {{Lei}}},\ }\bibfield  {title} {\enquote
  {\bibinfo {title} {{Testing the Einstein's equivalence principle with
  polarized gamma-ray bursts}},}\ }\href {\doibase 10.1093/mnrasl/slx045}
  {\bibfield  {journal} {\bibinfo  {journal} {Mon. Not. R. Astron. Soc. Lett.}\
  }\textbf {\bibinfo {volume} {469}},\ \bibinfo {pages} {L36} (\bibinfo {year}
  {2017})},\ \Eprint {http://arxiv.org/abs/1706.00889} {arXiv:1706.00889
  [astro-ph.HE]} \BibitemShut {NoStop}%
\bibitem [{\citenamefont {{Yu}}\ \emph {et~al.}(2018)\citenamefont {{Yu}},
  \citenamefont {{Xi}},\ and\ \citenamefont {{Wang}}}]{yu:2018ap}%
  \BibitemOpen
  \bibfield  {author} {\bibinfo {author} {\bibfnamefont {H.}~\bibnamefont
  {{Yu}}}, \bibinfo {author} {\bibfnamefont {S.-Q.}\ \bibnamefont {{Xi}}}, \
  and\ \bibinfo {author} {\bibfnamefont {F.-Y.}\ \bibnamefont {{Wang}}},\
  }\bibfield  {title} {\enquote {\bibinfo {title} {{A New Method to Test the
  Einstein's Weak Equivalence Principle}},}\ }\href {\doibase
  10.3847/1538-4357/aac2e3} {\bibfield  {journal} {\bibinfo  {journal}
  {Astrophys. J.}\ }\textbf {\bibinfo {volume} {860}},\ \bibinfo {eid} {173}
  (\bibinfo {year} {2018})},\ \Eprint {http://arxiv.org/abs/1708.02396}
  {arXiv:1708.02396 [astro-ph.HE]} \BibitemShut {NoStop}%
\bibitem [{\citenamefont {{Wei}}\ \emph {et~al.}(2017)\citenamefont {{Wei}},
  \citenamefont {{Zhang}}, \citenamefont {{Wu}}, \citenamefont {{Gao}},
  \citenamefont {{M{\'e}sz{\'a}ros}}, \citenamefont {{Zhang}}, \citenamefont
  {{Dai}}, \citenamefont {{Zhang}},\ and\ \citenamefont {{Zhu}}}]{wei:2017jc}%
  \BibitemOpen
  \bibfield  {author} {\bibinfo {author} {\bibfnamefont {J.-J.}\ \bibnamefont
  {{Wei}}}, \bibinfo {author} {\bibfnamefont {B.-B.}\ \bibnamefont {{Zhang}}},
  \bibinfo {author} {\bibfnamefont {X.-F.}\ \bibnamefont {{Wu}}}, \bibinfo
  {author} {\bibfnamefont {H.}~\bibnamefont {{Gao}}}, \bibinfo {author}
  {\bibfnamefont {P.}~\bibnamefont {{M{\'e}sz{\'a}ros}}}, \bibinfo {author}
  {\bibfnamefont {B.}~\bibnamefont {{Zhang}}}, \bibinfo {author} {\bibfnamefont
  {Z.-G.}\ \bibnamefont {{Dai}}}, \bibinfo {author} {\bibfnamefont {S.-N.}\
  \bibnamefont {{Zhang}}}, \ and\ \bibinfo {author} {\bibfnamefont {Z.-H.}\
  \bibnamefont {{Zhu}}},\ }\bibfield  {title} {\enquote {\bibinfo {title}
  {{Multimessenger tests of the weak equivalence principle from GW170817 and
  its electromagnetic counterparts}},}\ }\href {\doibase
  10.1088/1475-7516/2017/11/035} {\bibfield  {journal} {\bibinfo  {journal} {J.
  Cosmol. Astropart. Phys.}\ }\textbf {\bibinfo {volume} {11}},\ \bibinfo {eid}
  {035} (\bibinfo {year} {2017})},\ \Eprint {http://arxiv.org/abs/1710.05860}
  {arXiv:1710.05860 [astro-ph.HE]} \BibitemShut {NoStop}%
\bibitem [{\citenamefont {{Wang}}\ \emph {et~al.}(2017)\citenamefont {{Wang}},
  \citenamefont {{Zhang}}, \citenamefont {{Wang}}, \citenamefont {{Shen}},
  \citenamefont {{Liang}}, \citenamefont {{Li}}, \citenamefont {{Liao}},
  \citenamefont {{Jin}}, \citenamefont {{Yuan}}, \citenamefont {{Zou}},
  \citenamefont {{Fan}},\ and\ \citenamefont {{Wei}}}]{wang:2017ap}%
  \BibitemOpen
  \bibfield  {author} {\bibinfo {author} {\bibfnamefont {H.}~\bibnamefont
  {{Wang}}}, \bibinfo {author} {\bibfnamefont {F.-W.}\ \bibnamefont {{Zhang}}},
  \bibinfo {author} {\bibfnamefont {Y.-Z.}\ \bibnamefont {{Wang}}}, \bibinfo
  {author} {\bibfnamefont {Z.-Q.}\ \bibnamefont {{Shen}}}, \bibinfo {author}
  {\bibfnamefont {Y.-F.}\ \bibnamefont {{Liang}}}, \bibinfo {author}
  {\bibfnamefont {X.}~\bibnamefont {{Li}}}, \bibinfo {author} {\bibfnamefont
  {N.-H.}\ \bibnamefont {{Liao}}}, \bibinfo {author} {\bibfnamefont {Z.-P.}\
  \bibnamefont {{Jin}}}, \bibinfo {author} {\bibfnamefont {Q.}~\bibnamefont
  {{Yuan}}}, \bibinfo {author} {\bibfnamefont {Y.-C.}\ \bibnamefont {{Zou}}},
  \bibinfo {author} {\bibfnamefont {Y.-Z.}\ \bibnamefont {{Fan}}}, \ and\
  \bibinfo {author} {\bibfnamefont {D.-M.}\ \bibnamefont {{Wei}}},\ }\bibfield
  {title} {\enquote {\bibinfo {title} {{The GW170817/GRB 170817A/AT 2017gfo
  Association: Some Implications for Physics and Astrophysics}},}\ }\href
  {\doibase 10.3847/2041-8213/aa9e08} {\bibfield  {journal} {\bibinfo
  {journal} {Astrophys. J. Lett.}\ }\textbf {\bibinfo {volume} {851}},\
  \bibinfo {eid} {L18} (\bibinfo {year} {2017})},\ \Eprint
  {http://arxiv.org/abs/1710.05805} {arXiv:1710.05805 [astro-ph.HE]}
  \BibitemShut {NoStop}%
\bibitem [{\citenamefont {{Boran}}\ \emph {et~al.}(2018)\citenamefont
  {{Boran}}, \citenamefont {{Desai}}, \citenamefont {{Kahya}},\ and\
  \citenamefont {{Woodard}}}]{boran:2018pr}%
  \BibitemOpen
  \bibfield  {author} {\bibinfo {author} {\bibfnamefont {S.}~\bibnamefont
  {{Boran}}}, \bibinfo {author} {\bibfnamefont {S.}~\bibnamefont {{Desai}}},
  \bibinfo {author} {\bibfnamefont {E.~O.}\ \bibnamefont {{Kahya}}}, \ and\
  \bibinfo {author} {\bibfnamefont {R.~P.}\ \bibnamefont {{Woodard}}},\
  }\bibfield  {title} {\enquote {\bibinfo {title} {{GW170817 falsifies dark
  matter emulators}},}\ }\href {\doibase 10.1103/PhysRevD.97.041501} {\bibfield
   {journal} {\bibinfo  {journal} {\prd}\ }\textbf {\bibinfo {volume} {97}},\
  \bibinfo {eid} {041501(R)} (\bibinfo {year} {2018})},\ \Eprint
  {http://arxiv.org/abs/1710.06168} {arXiv:1710.06168 [astro-ph.HE]}
  \BibitemShut {NoStop}%
\bibitem [{\citenamefont {{Shoemaker}}\ and\ \citenamefont
  {{Murase}}(2018)}]{shoemaker:2018pr}%
  \BibitemOpen
  \bibfield  {author} {\bibinfo {author} {\bibfnamefont {I.~M.}\ \bibnamefont
  {{Shoemaker}}}\ and\ \bibinfo {author} {\bibfnamefont {K.}~\bibnamefont
  {{Murase}}},\ }\bibfield  {title} {\enquote {\bibinfo {title} {{Constraints
  from the time lag between gravitational waves and gamma rays: Implications of
  GW170817 and GRB 170817A}},}\ }\href {\doibase 10.1103/PhysRevD.97.083013}
  {\bibfield  {journal} {\bibinfo  {journal} {\prd}\ }\textbf {\bibinfo
  {volume} {97}},\ \bibinfo {eid} {083013} (\bibinfo {year} {2018})},\ \Eprint
  {http://arxiv.org/abs/1710.06427} {arXiv:1710.06427 [astro-ph.HE]}
  \BibitemShut {NoStop}%
\bibitem [{\citenamefont {{Bertolami}}\ and\ \citenamefont
  {{Landim}}(2018)}]{bertolami:2018pd}%
  \BibitemOpen
  \bibfield  {author} {\bibinfo {author} {\bibfnamefont {O.}~\bibnamefont
  {{Bertolami}}}\ and\ \bibinfo {author} {\bibfnamefont {R.~G.}\ \bibnamefont
  {{Landim}}},\ }\bibfield  {title} {\enquote {\bibinfo {title} {{Cosmic
  transients, Einstein's Equivalence Principle and dark matter halos}},}\
  }\href {\doibase 10.1016/j.dark.2018.05.002} {\bibfield  {journal} {\bibinfo
  {journal} {Phys. Dark Univ.}\ }\textbf {\bibinfo {volume} {21}},\ \bibinfo
  {pages} {16} (\bibinfo {year} {2018})},\ \Eprint
  {http://arxiv.org/abs/1712.04226} {arXiv:1712.04226 [gr-qc]} \BibitemShut
  {NoStop}%
\bibitem [{\citenamefont {{Leung}}\ \emph {et~al.}(2018)\citenamefont
  {{Leung}}, \citenamefont {{Hu}}, \citenamefont {{Harris}}, \citenamefont
  {{Brown}}, \citenamefont {{Gallicchio}},\ and\ \citenamefont
  {{Nguyen}}}]{leung:2018aj}%
  \BibitemOpen
  \bibfield  {author} {\bibinfo {author} {\bibfnamefont {C.}~\bibnamefont
  {{Leung}}}, \bibinfo {author} {\bibfnamefont {B.}~\bibnamefont {{Hu}}},
  \bibinfo {author} {\bibfnamefont {S.}~\bibnamefont {{Harris}}}, \bibinfo
  {author} {\bibfnamefont {A.}~\bibnamefont {{Brown}}}, \bibinfo {author}
  {\bibfnamefont {J.}~\bibnamefont {{Gallicchio}}}, \ and\ \bibinfo {author}
  {\bibfnamefont {H.}~\bibnamefont {{Nguyen}}},\ }\bibfield  {title} {\enquote
  {\bibinfo {title} {{Testing the Weak Equivalence Principle Using Optical and
  Near-infrared Crab Pulses}},}\ }\href {\doibase 10.3847/1538-4357/aac954}
  {\bibfield  {journal} {\bibinfo  {journal} {Astrophys. J.}\ }\textbf
  {\bibinfo {volume} {861}},\ \bibinfo {eid} {66} (\bibinfo {year} {2018})},\
  \Eprint {http://arxiv.org/abs/1804.04722} {arXiv:1804.04722 [astro-ph.HE]}
  \BibitemShut {NoStop}%
\bibitem [{\citenamefont {Boran}\ \emph {et~al.}(2019)\citenamefont {Boran},
  \citenamefont {Desai},\ and\ \citenamefont {Kahya}}]{Boran:2018ypz}%
  \BibitemOpen
  \bibfield  {author} {\bibinfo {author} {\bibfnamefont {S.}~\bibnamefont
  {Boran}}, \bibinfo {author} {\bibfnamefont {S.}~\bibnamefont {Desai}}, \ and\
  \bibinfo {author} {\bibfnamefont {E.~O.}\ \bibnamefont {Kahya}},\ }\bibfield
  {title} {\enquote {\bibinfo {title} {{Constraints on differential Shapiro
  delay between neutrinos and photons from IceCube-170922A}},}\ }\href
  {\doibase 10.1140/epjc/s10052-019-6695-6} {\bibfield  {journal} {\bibinfo
  {journal} {Eur. Phys. J. C}\ }\textbf {\bibinfo {volume} {79}},\ \bibinfo
  {pages} {185} (\bibinfo {year} {2019})},\ \Eprint
  {http://arxiv.org/abs/1807.05201} {arXiv:1807.05201 [astro-ph.HE]}
  \BibitemShut {NoStop}%
\bibitem [{\citenamefont {Laha}(2018)}]{Laha:2018hsh}%
  \BibitemOpen
  \bibfield  {author} {\bibinfo {author} {\bibfnamefont {R.}~\bibnamefont
  {Laha}},\ }\bibfield  {title} {\enquote {\bibinfo {title} {{Constraints on
  neutrino speed, weak equivalence principle violation, Lorentz invariance
  violation, and dual lensing from the first high-energy astrophysical neutrino
  source TXS 0506+056}},}\ }\href@noop {} {\  (\bibinfo {year} {2018})},\
  \Eprint {http://arxiv.org/abs/1807.05621} {arXiv:1807.05621 [astro-ph.HE]}
  \BibitemShut {NoStop}%
\bibitem [{\citenamefont {{Wei}}\ \emph {et~al.}(2019)\citenamefont {{Wei}},
  \citenamefont {{Zhang}}, \citenamefont {{Shao}}, \citenamefont {{Gao}},
  \citenamefont {{Li}}, \citenamefont {{Yin}}, \citenamefont {{Wu}},
  \citenamefont {{Wang}}, \citenamefont {{Zhang}},\ and\ \citenamefont
  {{Dai}}}]{wei:2019jh}%
  \BibitemOpen
  \bibfield  {author} {\bibinfo {author} {\bibfnamefont {J.-J.}\ \bibnamefont
  {{Wei}}}, \bibinfo {author} {\bibfnamefont {B.-B.}\ \bibnamefont {{Zhang}}},
  \bibinfo {author} {\bibfnamefont {L.}~\bibnamefont {{Shao}}}, \bibinfo
  {author} {\bibfnamefont {H.}~\bibnamefont {{Gao}}}, \bibinfo {author}
  {\bibfnamefont {Y.}~\bibnamefont {{Li}}}, \bibinfo {author} {\bibfnamefont
  {Q.-Q.}\ \bibnamefont {{Yin}}}, \bibinfo {author} {\bibfnamefont {X.-F.}\
  \bibnamefont {{Wu}}}, \bibinfo {author} {\bibfnamefont {X.-Y.}\ \bibnamefont
  {{Wang}}}, \bibinfo {author} {\bibfnamefont {B.}~\bibnamefont {{Zhang}}}, \
  and\ \bibinfo {author} {\bibfnamefont {Z.-G.}\ \bibnamefont {{Dai}}},\
  }\bibfield  {title} {\enquote {\bibinfo {title} {{Multimessenger tests of
  Einstein's weak equivalence principle and Lorentz invariance with a
  high-energy neutrino from a flaring blazar}},}\ }\href {\doibase
  10.1016/j.jheap.2019.01.002} {\bibfield  {journal} {\bibinfo  {journal} {J.
  High Energy Astrophys.}\ }\textbf {\bibinfo {volume} {22}},\ \bibinfo {pages}
  {1} (\bibinfo {year} {2019})},\ \Eprint {http://arxiv.org/abs/1807.06504}
  {arXiv:1807.06504 [astro-ph.HE]} \BibitemShut {NoStop}%
\bibitem [{\citenamefont {{Wei}}\ and\ \citenamefont
  {{Wu}}(2019)}]{wei:2019pr}%
  \BibitemOpen
  \bibfield  {author} {\bibinfo {author} {\bibfnamefont {J.-J.}\ \bibnamefont
  {{Wei}}}\ and\ \bibinfo {author} {\bibfnamefont {X.-F.}\ \bibnamefont
  {{Wu}}},\ }\bibfield  {title} {\enquote {\bibinfo {title} {{Precision test of
  the weak equivalence principle from gamma-ray burst polarization}},}\ }\href
  {\doibase 10.1103/PhysRevD.99.103012} {\bibfield  {journal} {\bibinfo
  {journal} {\prd}\ }\textbf {\bibinfo {volume} {99}},\ \bibinfo {eid} {103012}
  (\bibinfo {year} {2019})},\ \Eprint {http://arxiv.org/abs/1905.09995}
  {arXiv:1905.09995 [astro-ph.HE]} \BibitemShut {NoStop}%
\bibitem [{\citenamefont {Xing}\ \emph {et~al.}(2019)\citenamefont {Xing},
  \citenamefont {Gao}, \citenamefont {Wei}, \citenamefont {Li}, \citenamefont
  {Wang}, \citenamefont {Zhang}, \citenamefont {Wu},\ and\ \citenamefont
  {M{\'e}sz{\'a}ros}}]{Xing:2019geq}%
  \BibitemOpen
  \bibfield  {author} {\bibinfo {author} {\bibfnamefont {N.}~\bibnamefont
  {Xing}}, \bibinfo {author} {\bibfnamefont {H.}~\bibnamefont {Gao}}, \bibinfo
  {author} {\bibfnamefont {J.}~\bibnamefont {Wei}}, \bibinfo {author}
  {\bibfnamefont {Z.}~\bibnamefont {Li}}, \bibinfo {author} {\bibfnamefont
  {W.}~\bibnamefont {Wang}}, \bibinfo {author} {\bibfnamefont {B.}~\bibnamefont
  {Zhang}}, \bibinfo {author} {\bibfnamefont {X.-F.}\ \bibnamefont {Wu}}, \
  and\ \bibinfo {author} {\bibfnamefont {P.}~\bibnamefont {M{\'e}sz{\'a}ros}},\
  }\bibfield  {title} {\enquote {\bibinfo {title} {{Limits on the Weak
  Equivalence Principle and Photon Mass with FRB 121102 Subpulses}},}\ }\href
  {\doibase 10.3847/2041-8213/ab3c5f} {\bibfield  {journal} {\bibinfo
  {journal} {Astrophys. J. Lett.}\ }\textbf {\bibinfo {volume} {882}},\
  \bibinfo {pages} {L13} (\bibinfo {year} {2019})},\ \Eprint
  {http://arxiv.org/abs/1907.00583} {arXiv:1907.00583 [astro-ph.HE]}
  \BibitemShut {NoStop}%
\bibitem [{\citenamefont {Yao}\ \emph {et~al.}(2019)\citenamefont {Yao},
  \citenamefont {Zhao}, \citenamefont {Han}, \citenamefont {Wang},
  \citenamefont {Liu},\ and\ \citenamefont {Liu}}]{Yao:2019cku}%
  \BibitemOpen
  \bibfield  {author} {\bibinfo {author} {\bibfnamefont {L.}~\bibnamefont
  {Yao}}, \bibinfo {author} {\bibfnamefont {Z.}~\bibnamefont {Zhao}}, \bibinfo
  {author} {\bibfnamefont {Y.}~\bibnamefont {Han}}, \bibinfo {author}
  {\bibfnamefont {J.}~\bibnamefont {Wang}}, \bibinfo {author} {\bibfnamefont
  {T.}~\bibnamefont {Liu}}, \ and\ \bibinfo {author} {\bibfnamefont
  {M.}~\bibnamefont {Liu}},\ }\bibfield  {title} {\enquote {\bibinfo {title}
  {{A new test of the weak equivalence principle on the galaxy level based on
  observations of binary neutron star merger GW170817}},}\ }\href@noop {} {\
  (\bibinfo {year} {2019})},\ \Eprint {http://arxiv.org/abs/1909.04338}
  {arXiv:1909.04338 [astro-ph.HE]} \BibitemShut {NoStop}%
\bibitem [{\citenamefont {{Longo}}(1988)}]{longo:1988pl}%
  \BibitemOpen
  \bibfield  {author} {\bibinfo {author} {\bibfnamefont {M.~J.}\ \bibnamefont
  {{Longo}}},\ }\bibfield  {title} {\enquote {\bibinfo {title} {{New precision
  tests of the Einstein equivalence principle from SN1987A}},}\ }\href
  {\doibase 10.1103/PhysRevLett.60.173} {\bibfield  {journal} {\bibinfo
  {journal} {Phys. Rev. Lett.}\ }\textbf {\bibinfo {volume} {60}},\ \bibinfo
  {pages} {173} (\bibinfo {year} {1988})}\BibitemShut {NoStop}%
\bibitem [{\citenamefont {{Krauss}}\ and\ \citenamefont
  {{Tremaine}}(1988)}]{krauss:1988pl}%
  \BibitemOpen
  \bibfield  {author} {\bibinfo {author} {\bibfnamefont {L.~M.}\ \bibnamefont
  {{Krauss}}}\ and\ \bibinfo {author} {\bibfnamefont {S.}~\bibnamefont
  {{Tremaine}}},\ }\bibfield  {title} {\enquote {\bibinfo {title} {{Test of the
  weak equivalence principle for neutrinos and photons}},}\ }\href {\doibase
  10.1103/PhysRevLett.60.176} {\bibfield  {journal} {\bibinfo  {journal} {Phys.
  Rev. Lett.}\ }\textbf {\bibinfo {volume} {60}},\ \bibinfo {pages} {176}
  (\bibinfo {year} {1988})}\BibitemShut {NoStop}%
\bibitem [{\citenamefont {LoSecco}(1988)}]{LoSecco:1988jg}%
  \BibitemOpen
  \bibfield  {author} {\bibinfo {author} {\bibfnamefont {J.~M.}\ \bibnamefont
  {LoSecco}},\ }\bibfield  {title} {\enquote {\bibinfo {title} {{Limits on {CP}
  Invariance in General Relativity}},}\ }\href {\doibase
  10.1103/PhysRevD.38.3313} {\bibfield  {journal} {\bibinfo  {journal} {Phys.
  Rev. D}\ }\textbf {\bibinfo {volume} {38}},\ \bibinfo {pages} {3313}
  (\bibinfo {year} {1988})}\BibitemShut {NoStop}%
\bibitem [{\citenamefont {{Pakvasa}}\ \emph {et~al.}(1989)\citenamefont
  {{Pakvasa}}, \citenamefont {{Simmons}},\ and\ \citenamefont
  {{Weiler}}}]{pakvasa:1989pr}%
  \BibitemOpen
  \bibfield  {author} {\bibinfo {author} {\bibfnamefont {S.}~\bibnamefont
  {{Pakvasa}}}, \bibinfo {author} {\bibfnamefont {W.~A.}\ \bibnamefont
  {{Simmons}}}, \ and\ \bibinfo {author} {\bibfnamefont {T.~J.}\ \bibnamefont
  {{Weiler}}},\ }\bibfield  {title} {\enquote {\bibinfo {title} {{Test of
  equivalence principle for neutrinos and antineutrinos}},}\ }\href {\doibase
  10.1103/PhysRevD.39.1761} {\bibfield  {journal} {\bibinfo  {journal} {\prd}\
  }\textbf {\bibinfo {volume} {39}},\ \bibinfo {pages} {1761} (\bibinfo {year}
  {1989})}\BibitemShut {NoStop}%
\bibitem [{\citenamefont {Bose}\ and\ \citenamefont
  {McGlinn}(1988)}]{Bose:1988wi}%
  \BibitemOpen
  \bibfield  {author} {\bibinfo {author} {\bibfnamefont {S.~K.}\ \bibnamefont
  {Bose}}\ and\ \bibinfo {author} {\bibfnamefont {W.~D.}\ \bibnamefont
  {McGlinn}},\ }\bibfield  {title} {\enquote {\bibinfo {title} {{EFFECT OF
  FINITE MASS ON GRAVITATIONAL TRANSIT TIME}},}\ }\href {\doibase
  10.1103/PhysRevD.38.2335} {\bibfield  {journal} {\bibinfo  {journal} {Phys.
  Rev. D}\ }\textbf {\bibinfo {volume} {38}},\ \bibinfo {pages} {2335}
  (\bibinfo {year} {1988})}\BibitemShut {NoStop}%
\bibitem [{\citenamefont {Shapiro}(1964)}]{Shapiro:1964uw}%
  \BibitemOpen
  \bibfield  {author} {\bibinfo {author} {\bibfnamefont {I.~I.}\ \bibnamefont
  {Shapiro}},\ }\bibfield  {title} {\enquote {\bibinfo {title} {{Fourth Test of
  General Relativity}},}\ }\href {\doibase 10.1103/PhysRevLett.13.789}
  {\bibfield  {journal} {\bibinfo  {journal} {Phys. Rev. Lett.}\ }\textbf
  {\bibinfo {volume} {13}},\ \bibinfo {pages} {789} (\bibinfo {year}
  {1964})}\BibitemShut {NoStop}%
\bibitem [{\citenamefont {{Blandford}}\ and\ \citenamefont
  {{Narayan}}(1986)}]{blandford:1986aj}%
  \BibitemOpen
  \bibfield  {author} {\bibinfo {author} {\bibfnamefont {R.}~\bibnamefont
  {{Blandford}}}\ and\ \bibinfo {author} {\bibfnamefont {R.}~\bibnamefont
  {{Narayan}}},\ }\bibfield  {title} {\enquote {\bibinfo {title} {{Fermat's
  Principle, Caustics, and the Classification of Gravitational Lens Images}},}\
  }\href {\doibase 10.1086/164709} {\bibfield  {journal} {\bibinfo  {journal}
  {Astrophys. J.}\ }\textbf {\bibinfo {volume} {310}},\ \bibinfo {pages} {568}
  (\bibinfo {year} {1986})}\BibitemShut {NoStop}%
\bibitem [{\citenamefont {{Wambsganss}}(1998)}]{wambsganss:1998lr}%
  \BibitemOpen
  \bibfield  {author} {\bibinfo {author} {\bibfnamefont {J.}~\bibnamefont
  {{Wambsganss}}},\ }\bibfield  {title} {\enquote {\bibinfo {title}
  {{Gravitational Lensing in Astronomy}},}\ }\href {\doibase
  10.12942/lrr-1998-12} {\bibfield  {journal} {\bibinfo  {journal} {Living Rev.
  Relativity}\ }\textbf {\bibinfo {volume} {1}},\ \bibinfo {eid} {12} (\bibinfo
  {year} {1998})},\ \Eprint {http://arxiv.org/abs/astro-ph/9812021}
  {arXiv:astro-ph/9812021 [astro-ph]} \BibitemShut {NoStop}%
\bibitem [{\citenamefont {{Abbott}}\ \emph
  {et~al.}(2017{\natexlab{a}})\citenamefont {{Abbott}} \emph
  {et~al.}}]{abbott:2017aj}%
  \BibitemOpen
  \bibfield  {author} {\bibinfo {author} {\bibfnamefont {B.~P.}\ \bibnamefont
  {{Abbott}}} \emph {et~al.} (\bibinfo {collaboration} {LIGO Scientific
  Collaboration, Virgo Collaboration, Fermi GBM, and INTEGRAL}),\ }\bibfield
  {title} {\enquote {\bibinfo {title} {{Gravitational Waves and Gamma-Rays from
  a Binary Neutron Star Merger: GW170817 and GRB 170817A}},}\ }\href {\doibase
  10.3847/2041-8213/aa920c} {\bibfield  {journal} {\bibinfo  {journal}
  {Astrophys. J. Lett.}\ }\textbf {\bibinfo {volume} {848}},\ \bibinfo {eid}
  {L13} (\bibinfo {year} {2017}{\natexlab{a}})},\ \Eprint
  {http://arxiv.org/abs/1710.05834} {arXiv:1710.05834 [astro-ph.HE]}
  \BibitemShut {NoStop}%
\bibitem [{\citenamefont {{Teyssandier}}\ and\ \citenamefont {{Le
  Poncin-Lafitte}}(2008)}]{teyssandier2008cq}%
  \BibitemOpen
  \bibfield  {author} {\bibinfo {author} {\bibfnamefont {P.}~\bibnamefont
  {{Teyssandier}}}\ and\ \bibinfo {author} {\bibfnamefont {C.}~\bibnamefont
  {{Le Poncin-Lafitte}}},\ }\bibfield  {title} {\enquote {\bibinfo {title}
  {{General post-Minkowskian expansion of time transfer functions}},}\ }\href
  {\doibase 10.1088/0264-9381/25/14/145020} {\bibfield  {journal} {\bibinfo
  {journal} {Classical Quantum Gravity}\ }\textbf {\bibinfo {volume} {25}},\
  \bibinfo {eid} {145020} (\bibinfo {year} {2008})},\ \Eprint
  {http://arxiv.org/abs/0803.0277} {arXiv:0803.0277 [gr-qc]} \BibitemShut
  {NoStop}%
\bibitem [{\citenamefont {{Gao}}\ and\ \citenamefont
  {{Wald}}(2000)}]{gao:2000cq}%
  \BibitemOpen
  \bibfield  {author} {\bibinfo {author} {\bibfnamefont {S.}~\bibnamefont
  {{Gao}}}\ and\ \bibinfo {author} {\bibfnamefont {R.~M.}\ \bibnamefont
  {{Wald}}},\ }\bibfield  {title} {\enquote {\bibinfo {title} {{Theorems on
  gravitational time delay and related issues}},}\ }\href {\doibase
  10.1088/0264-9381/17/24/305} {\bibfield  {journal} {\bibinfo  {journal}
  {Classical Quantum Gravity}\ }\textbf {\bibinfo {volume} {17}},\ \bibinfo
  {pages} {4999} (\bibinfo {year} {2000})},\ \Eprint
  {http://arxiv.org/abs/gr-qc/0007021} {arXiv:gr-qc/0007021 [gr-qc]}
  \BibitemShut {NoStop}%
\bibitem [{\citenamefont {{Sanghai}}\ and\ \citenamefont
  {{Clifton}}(2016)}]{sanghai:2016}%
  \BibitemOpen
  \bibfield  {author} {\bibinfo {author} {\bibfnamefont {V.~A.~A.}\
  \bibnamefont {{Sanghai}}}\ and\ \bibinfo {author} {\bibfnamefont
  {T.}~\bibnamefont {{Clifton}}},\ }\bibfield  {title} {\enquote {\bibinfo
  {title} {{Cosmological backreaction in the presence of radiation and a
  cosmological constant}},}\ }\href {\doibase 10.1103/PhysRevD.94.023505}
  {\bibfield  {journal} {\bibinfo  {journal} {\prd}\ }\textbf {\bibinfo
  {volume} {94}},\ \bibinfo {eid} {023505} (\bibinfo {year} {2016})},\ \Eprint
  {http://arxiv.org/abs/1604.06345} {arXiv:1604.06345 [gr-qc]} \BibitemShut
  {NoStop}%
\bibitem [{\citenamefont {Will}(2014)}]{lrr-2014-4}%
  \BibitemOpen
  \bibfield  {author} {\bibinfo {author} {\bibfnamefont {C.~M.}\ \bibnamefont
  {Will}},\ }\bibfield  {title} {\enquote {\bibinfo {title} {{The Confrontation
  between General Relativity and Experiment}},}\ }\href {\doibase
  10.12942/lrr-2014-4} {\bibfield  {journal} {\bibinfo  {journal} {Living Rev.
  Relativity}\ }\textbf {\bibinfo {volume} {17}},\ \bibinfo {pages} {4}
  (\bibinfo {year} {2014})},\ \Eprint {http://arxiv.org/abs/1403.7377}
  {arXiv:1403.7377 [gr-qc]} \BibitemShut {NoStop}%
\bibitem [{\citenamefont {{Will}}(1985)}]{will:1985bk}%
  \BibitemOpen
  \bibfield  {author} {\bibinfo {author} {\bibfnamefont {C.~M.}\ \bibnamefont
  {{Will}}},\ }\href@noop {} {\emph {\bibinfo {title} {{Theory and experiment
  in gravitational physics.}}}}\ (\bibinfo  {publisher} {Cambridge University
  Press},\ \bibinfo {year} {1985})\BibitemShut {NoStop}%
\bibitem [{\citenamefont {Perlick}(2004)}]{Perlick:2004tq}%
  \BibitemOpen
  \bibfield  {author} {\bibinfo {author} {\bibfnamefont {V.}~\bibnamefont
  {Perlick}},\ }\bibfield  {title} {\enquote {\bibinfo {title} {{Gravitational
  lensing from a spacetime perspective}},}\ }\href {\doibase
  10.12942/lrr-2004-9} {\bibfield  {journal} {\bibinfo  {journal} {Living Rev.
  Relativity}\ }\textbf {\bibinfo {volume} {7}},\ \bibinfo {pages} {9}
  (\bibinfo {year} {2004})}\BibitemShut {NoStop}%
\bibitem [{\citenamefont {{Etherington}}(1933)}]{etherington:1933rm}%
  \BibitemOpen
  \bibfield  {author} {\bibinfo {author} {\bibfnamefont {I.~M.~H.}\
  \bibnamefont {{Etherington}}},\ }\bibfield  {title} {\enquote {\bibinfo
  {title} {{On the Definition of Distance in General Relativity.}}}\ }\href
  {\doibase 10.1080/14786443309462220} {\bibfield  {journal} {\bibinfo
  {journal} {Philos. Mag.}\ }\textbf {\bibinfo {volume} {15}},\ \bibinfo
  {pages} {761} (\bibinfo {year} {1933})}\BibitemShut {NoStop}%
\bibitem [{\citenamefont {{Etherington}}(2007)}]{etherington:2007sf}%
  \BibitemOpen
  \bibfield  {author} {\bibinfo {author} {\bibfnamefont {I.~M.~H.}\
  \bibnamefont {{Etherington}}},\ }\bibfield  {title} {\enquote {\bibinfo
  {title} {{Republication of: LX. On the definition of distance in general
  relativity}},}\ }\href {\doibase 10.1007/s10714-007-0447-x} {\bibfield
  {journal} {\bibinfo  {journal} {Gen. Rel. Grav.}\ }\textbf {\bibinfo {volume}
  {39}},\ \bibinfo {pages} {1055} (\bibinfo {year} {2007})}\BibitemShut
  {NoStop}%
\bibitem [{\citenamefont {{Ellis}}(2007)}]{ellis:2007vn}%
  \BibitemOpen
  \bibfield  {author} {\bibinfo {author} {\bibfnamefont {G.~F.~R.}\
  \bibnamefont {{Ellis}}},\ }\bibfield  {title} {\enquote {\bibinfo {title}
  {{On the definition of distance in general relativity: I. M. H. Etherington
  (Philosophical Magazine ser. 7, vol. 15, 761 (1933))}},}\ }\href {\doibase
  10.1007/s10714-006-0355-5} {\bibfield  {journal} {\bibinfo  {journal} {Gen.
  Rel. Grav.}\ }\textbf {\bibinfo {volume} {39}},\ \bibinfo {pages} {1047}
  (\bibinfo {year} {2007})}\BibitemShut {NoStop}%
\bibitem [{\citenamefont {{Ellis}}(2009)}]{ellis:2009fk}%
  \BibitemOpen
  \bibfield  {author} {\bibinfo {author} {\bibfnamefont {G.~F.~R.}\
  \bibnamefont {{Ellis}}},\ }\bibfield  {title} {\enquote {\bibinfo {title}
  {{Republication of: Relativistic cosmology}},}\ }\href {\doibase
  10.1007/s10714-009-0760-7} {\bibfield  {journal} {\bibinfo  {journal} {Gen.
  Rel. Grav.}\ }\textbf {\bibinfo {volume} {41}},\ \bibinfo {pages} {581}
  (\bibinfo {year} {2009})}\BibitemShut {NoStop}%
\bibitem [{\citenamefont {{Ellis}}\ \emph {et~al.}(2013)\citenamefont
  {{Ellis}}, \citenamefont {{Poltis}}, \citenamefont {{Uzan}},\ and\
  \citenamefont {{Weltman}}}]{ellis:2013fk}%
  \BibitemOpen
  \bibfield  {author} {\bibinfo {author} {\bibfnamefont {G.~F.~R.}\
  \bibnamefont {{Ellis}}}, \bibinfo {author} {\bibfnamefont {R.}~\bibnamefont
  {{Poltis}}}, \bibinfo {author} {\bibfnamefont {J.-P.}\ \bibnamefont
  {{Uzan}}}, \ and\ \bibinfo {author} {\bibfnamefont {A.}~\bibnamefont
  {{Weltman}}},\ }\bibfield  {title} {\enquote {\bibinfo {title} {{Blackness of
  the cosmic microwave background spectrum as a probe of the distance-duality
  relation}},}\ }\href {\doibase 10.1103/PhysRevD.87.103530} {\bibfield
  {journal} {\bibinfo  {journal} {\prd}\ }\textbf {\bibinfo {volume} {87}},\
  \bibinfo {eid} {103530} (\bibinfo {year} {2013})},\ \Eprint
  {http://arxiv.org/abs/1301.1312} {arXiv:1301.1312 [astro-ph.CO]} \BibitemShut
  {NoStop}%
\bibitem [{\citenamefont {{Avni}}\ and\ \citenamefont
  {{Shulami}}(1988)}]{1988ApJ...332..113A}%
  \BibitemOpen
  \bibfield  {author} {\bibinfo {author} {\bibfnamefont {Y.}~\bibnamefont
  {{Avni}}}\ and\ \bibinfo {author} {\bibfnamefont {I.}~\bibnamefont
  {{Shulami}}},\ }\bibfield  {title} {\enquote {\bibinfo {title} {{``Flux
  Conservations'' by a Schwarzschild Gravitational Lens}},}\ }\href {\doibase
  10.1086/166636} {\bibfield  {journal} {\bibinfo  {journal} {Astrophys. J.}\
  }\textbf {\bibinfo {volume} {332}},\ \bibinfo {pages} {113} (\bibinfo {year}
  {1988})}\BibitemShut {NoStop}%
\bibitem [{\citenamefont {{Coley}}\ and\ \citenamefont
  {{Tremaine}}(1988)}]{coley:1988pr}%
  \BibitemOpen
  \bibfield  {author} {\bibinfo {author} {\bibfnamefont {A.~A.}\ \bibnamefont
  {{Coley}}}\ and\ \bibinfo {author} {\bibfnamefont {S.}~\bibnamefont
  {{Tremaine}}},\ }\bibfield  {title} {\enquote {\bibinfo {title} {{Constraint
  on nonmetric theories of gravity from supernova 1987A}},}\ }\href {\doibase
  10.1103/PhysRevD.38.2927} {\bibfield  {journal} {\bibinfo  {journal} {\prd}\
  }\textbf {\bibinfo {volume} {38}},\ \bibinfo {pages} {2927} (\bibinfo {year}
  {1988})}\BibitemShut {NoStop}%
\bibitem [{\citenamefont {{Kahya}}\ and\ \citenamefont
  {{Woodard}}(2007)}]{kahya:2007pl}%
  \BibitemOpen
  \bibfield  {author} {\bibinfo {author} {\bibfnamefont {E.~O.}\ \bibnamefont
  {{Kahya}}}\ and\ \bibinfo {author} {\bibfnamefont {R.~P.}\ \bibnamefont
  {{Woodard}}},\ }\bibfield  {title} {\enquote {\bibinfo {title} {{A generic
  test of modified gravity models which emulate dark matter}},}\ }\href
  {\doibase 10.1016/j.physletb.2007.07.029} {\bibfield  {journal} {\bibinfo
  {journal} {Phys. Lett. B}\ }\textbf {\bibinfo {volume} {652}},\ \bibinfo
  {pages} {213} (\bibinfo {year} {2007})},\ \Eprint
  {http://arxiv.org/abs/0705.0153} {arXiv:0705.0153} \BibitemShut {NoStop}%
\bibitem [{\citenamefont {Aghanim}\ \emph {et~al.}(2018)\citenamefont {Aghanim}
  \emph {et~al.}}]{Aghanim:2018eyx}%
  \BibitemOpen
  \bibfield  {author} {\bibinfo {author} {\bibfnamefont {N.}~\bibnamefont
  {Aghanim}} \emph {et~al.} (\bibinfo {collaboration} {Planck Collaboration}),\
  }\bibfield  {title} {\enquote {\bibinfo {title} {{Planck 2018 results. VI.
  Cosmological parameters}},}\ }\href@noop {} {\  (\bibinfo {year} {2018})},\
  \Eprint {http://arxiv.org/abs/1807.06209} {arXiv:1807.06209 [astro-ph.CO]}
  \BibitemShut {NoStop}%
\bibitem [{\citenamefont {Nicolis}\ \emph {et~al.}(2009)\citenamefont
  {Nicolis}, \citenamefont {Rattazzi},\ and\ \citenamefont
  {Trincherini}}]{Nicolis:2008in}%
  \BibitemOpen
  \bibfield  {author} {\bibinfo {author} {\bibfnamefont {A.}~\bibnamefont
  {Nicolis}}, \bibinfo {author} {\bibfnamefont {R.}~\bibnamefont {Rattazzi}}, \
  and\ \bibinfo {author} {\bibfnamefont {E.}~\bibnamefont {Trincherini}},\
  }\bibfield  {title} {\enquote {\bibinfo {title} {{Galileon as a local
  modification of gravity}},}\ }\href {\doibase 10.1103/PhysRevD.79.064036}
  {\bibfield  {journal} {\bibinfo  {journal} {Phys. Rev. D}\ }\textbf {\bibinfo
  {volume} {79}},\ \bibinfo {pages} {064036} (\bibinfo {year} {2009})},\
  \Eprint {http://arxiv.org/abs/0811.2197} {arXiv:0811.2197 [hep-th]}
  \BibitemShut {NoStop}%
\bibitem [{\citenamefont {Holz}\ and\ \citenamefont
  {Wald}(1998)}]{Holz:1997ic}%
  \BibitemOpen
  \bibfield  {author} {\bibinfo {author} {\bibfnamefont {D.~E.}\ \bibnamefont
  {Holz}}\ and\ \bibinfo {author} {\bibfnamefont {R.~M.}\ \bibnamefont
  {Wald}},\ }\bibfield  {title} {\enquote {\bibinfo {title} {{New method for
  determining cumulative gravitational lensing effects in inhomogeneous
  universes}},}\ }\href {\doibase 10.1103/PhysRevD.58.063501} {\bibfield
  {journal} {\bibinfo  {journal} {Phys. Rev. D}\ }\textbf {\bibinfo {volume}
  {58}},\ \bibinfo {pages} {063501} (\bibinfo {year} {1998})},\ \Eprint
  {http://arxiv.org/abs/astro-ph/9708036} {arXiv:astro-ph/9708036 [astro-ph]}
  \BibitemShut {NoStop}%
\bibitem [{\citenamefont {Abbott}\ \emph {et~al.}(2019)\citenamefont {Abbott}
  \emph {et~al.}}]{Abbott:2018wiz}%
  \BibitemOpen
  \bibfield  {author} {\bibinfo {author} {\bibfnamefont {B.~P.}\ \bibnamefont
  {Abbott}} \emph {et~al.} (\bibinfo {collaboration} {LIGO Scientific
  Collaboration and Virgo Collaboration}),\ }\bibfield  {title} {\enquote
  {\bibinfo {title} {{Properties of the binary neutron star merger
  GW170817}},}\ }\href {\doibase 10.1103/PhysRevX.9.011001} {\bibfield
  {journal} {\bibinfo  {journal} {Phys. Rev. X}\ }\textbf {\bibinfo {volume}
  {9}},\ \bibinfo {pages} {011001} (\bibinfo {year} {2019})},\ \Eprint
  {http://arxiv.org/abs/1805.11579} {arXiv:1805.11579 [gr-qc]} \BibitemShut
  {NoStop}%
\bibitem [{\citenamefont {McVittie}(1933)}]{McVittie:1933zz}%
  \BibitemOpen
  \bibfield  {author} {\bibinfo {author} {\bibfnamefont {G.~C.}\ \bibnamefont
  {McVittie}},\ }\bibfield  {title} {\enquote {\bibinfo {title} {{The
  mass-particle in an expanding universe}},}\ }\href {\doibase
  10.1093/mnras/93.5.325} {\bibfield  {journal} {\bibinfo  {journal} {Mon. Not.
  R. Astron. Soc.}\ }\textbf {\bibinfo {volume} {93}},\ \bibinfo {pages} {325}
  (\bibinfo {year} {1933})}\BibitemShut {NoStop}%
\bibitem [{\citenamefont {Shaw}\ and\ \citenamefont
  {Barrow}(2006)}]{Shaw:2005gt}%
  \BibitemOpen
  \bibfield  {author} {\bibinfo {author} {\bibfnamefont {D.~J.}\ \bibnamefont
  {Shaw}}\ and\ \bibinfo {author} {\bibfnamefont {J.~D.}\ \bibnamefont
  {Barrow}},\ }\bibfield  {title} {\enquote {\bibinfo {title} {{Local effects
  of cosmological variations in physical `constants' and scalar fields. I.
  Spherically symmetric spacetimes}},}\ }\href {\doibase
  10.1103/PhysRevD.73.123505} {\bibfield  {journal} {\bibinfo  {journal} {Phys.
  Rev. D}\ }\textbf {\bibinfo {volume} {73}},\ \bibinfo {pages} {123505}
  (\bibinfo {year} {2006})},\ \Eprint {http://arxiv.org/abs/gr-qc/0512022}
  {arXiv:gr-qc/0512022 [gr-qc]} \BibitemShut {NoStop}%
\bibitem [{\citenamefont {{Abbott}}\ \emph
  {et~al.}(2017{\natexlab{b}})\citenamefont {{Abbott}} \emph
  {et~al.}}]{abbott:2017pl}%
  \BibitemOpen
  \bibfield  {author} {\bibinfo {author} {\bibfnamefont {B.~P.}\ \bibnamefont
  {{Abbott}}} \emph {et~al.} (\bibinfo {collaboration} {LIGO Scientific
  Collaboration and Virgo Collaboration}),\ }\bibfield  {title} {\enquote
  {\bibinfo {title} {{GW170817: Observation of Gravitational Waves from a
  Binary Neutron Star Inspiral}},}\ }\href {\doibase
  10.1103/PhysRevLett.119.161101} {\bibfield  {journal} {\bibinfo  {journal}
  {Phys. Rev. Lett.}\ }\textbf {\bibinfo {volume} {119}},\ \bibinfo {eid}
  {161101} (\bibinfo {year} {2017}{\natexlab{b}})},\ \Eprint
  {http://arxiv.org/abs/1710.05832} {arXiv:1710.05832 [gr-qc]} \BibitemShut
  {NoStop}%
\bibitem [{\citenamefont {Green}\ and\ \citenamefont
  {Wald}(2014)}]{Green:2014aga}%
  \BibitemOpen
  \bibfield  {author} {\bibinfo {author} {\bibfnamefont {S.~R.}\ \bibnamefont
  {Green}}\ and\ \bibinfo {author} {\bibfnamefont {R.~M.}\ \bibnamefont
  {Wald}},\ }\bibfield  {title} {\enquote {\bibinfo {title} {{How well is our
  universe described by an FLRW model?}}}\ }\href {\doibase
  10.1088/0264-9381/31/23/234003} {\bibfield  {journal} {\bibinfo  {journal}
  {Classical Quantum Gravity}\ }\textbf {\bibinfo {volume} {31}},\ \bibinfo
  {pages} {234003} (\bibinfo {year} {2014})},\ \Eprint
  {http://arxiv.org/abs/1407.8084} {arXiv:1407.8084 [gr-qc]} \BibitemShut
  {NoStop}%
\bibitem [{\citenamefont {{Sanghai}}\ and\ \citenamefont
  {{Clifton}}(2015)}]{sanghai:2015}%
  \BibitemOpen
  \bibfield  {author} {\bibinfo {author} {\bibfnamefont {V.~A.~A.}\
  \bibnamefont {{Sanghai}}}\ and\ \bibinfo {author} {\bibfnamefont
  {T.}~\bibnamefont {{Clifton}}},\ }\bibfield  {title} {\enquote {\bibinfo
  {title} {{Post-Newtonian cosmological modelling}},}\ }\href {\doibase
  10.1103/PhysRevD.91.103532} {\bibfield  {journal} {\bibinfo  {journal}
  {\prd}\ }\textbf {\bibinfo {volume} {91}},\ \bibinfo {eid} {103532} (\bibinfo
  {year} {2015})},\ \Eprint {http://arxiv.org/abs/1503.08747} {arXiv:1503.08747
  [gr-qc]} \BibitemShut {NoStop}%
\bibitem [{\citenamefont {Glasser}\ and\ \citenamefont
  {Zucker}(1980)}]{GlasserZucker}%
  \BibitemOpen
  \bibfield  {author} {\bibinfo {author} {\bibfnamefont {M.~L.}\ \bibnamefont
  {Glasser}}\ and\ \bibinfo {author} {\bibfnamefont {I.~J.}\ \bibnamefont
  {Zucker}},\ }\bibfield  {title} {\enquote {\bibinfo {title} {Lattice sums},}\
  }in\ \href@noop {} {\emph {\bibinfo {booktitle} {Theoretical Chemistry:
  Advances and Perspectives}}},\ Vol.~\bibinfo {volume} {5},\ \bibinfo {editor}
  {edited by\ \bibinfo {editor} {\bibfnamefont {H.}~\bibnamefont {Eyring}}\
  and\ \bibinfo {editor} {\bibfnamefont {D.}~\bibnamefont {Henderson}}}\
  (\bibinfo  {publisher} {Academic Press},\ \bibinfo {address} {New York},\
  \bibinfo {year} {1980})\ pp.\ \bibinfo {pages} {67--139}\BibitemShut
  {NoStop}%
\bibitem [{\citenamefont {{Soffel}}\ \emph {et~al.}(2003)\citenamefont
  {{Soffel}} \emph {et~al.}}]{soffel:2003}%
  \BibitemOpen
  \bibfield  {author} {\bibinfo {author} {\bibfnamefont {M.}~\bibnamefont
  {{Soffel}}} \emph {et~al.},\ }\bibfield  {title} {\enquote {\bibinfo {title}
  {{The IAU 2000 Resolutions for Astrometry, Celestial Mechanics, and Metrology
  in the Relativistic Framework: Explanatory Supplement}},}\ }\href {\doibase
  10.1086/378162} {\bibfield  {journal} {\bibinfo  {journal} {Astrophys. J.}\
  }\textbf {\bibinfo {volume} {126}},\ \bibinfo {pages} {2687} (\bibinfo {year}
  {2003})},\ \Eprint {http://arxiv.org/abs/astro-ph/0303376} {astro-ph/0303376}
  \BibitemShut {NoStop}%
\bibitem [{\citenamefont {{Sanghai}}\ and\ \citenamefont
  {{Clifton}}(2017)}]{sanghai:2017}%
  \BibitemOpen
  \bibfield  {author} {\bibinfo {author} {\bibfnamefont {V.~A.~A.}\
  \bibnamefont {{Sanghai}}}\ and\ \bibinfo {author} {\bibfnamefont
  {T.}~\bibnamefont {{Clifton}}},\ }\bibfield  {title} {\enquote {\bibinfo
  {title} {{Parameterized post-Newtonian cosmology}},}\ }\href {\doibase
  10.1088/1361-6382/aa5d75} {\bibfield  {journal} {\bibinfo  {journal}
  {Classical Quantum Gravity}\ }\textbf {\bibinfo {volume} {34}},\ \bibinfo
  {eid} {065003} (\bibinfo {year} {2017})},\ \Eprint
  {http://arxiv.org/abs/1610.08039} {arXiv:1610.08039 [gr-qc]} \BibitemShut
  {NoStop}%
\bibitem [{\citenamefont {{Klioner}}\ and\ \citenamefont
  {{Soffel}}(2000)}]{klioner:2000pr}%
  \BibitemOpen
  \bibfield  {author} {\bibinfo {author} {\bibfnamefont {S.~A.}\ \bibnamefont
  {{Klioner}}}\ and\ \bibinfo {author} {\bibfnamefont {M.~H.}\ \bibnamefont
  {{Soffel}}},\ }\bibfield  {title} {\enquote {\bibinfo {title} {{Relativistic
  celestial mechanics with PPN parameters}},}\ }\href {\doibase
  10.1103/PhysRevD.62.024019} {\bibfield  {journal} {\bibinfo  {journal}
  {\prd}\ }\textbf {\bibinfo {volume} {62}},\ \bibinfo {eid} {024019} (\bibinfo
  {year} {2000})},\ \Eprint {http://arxiv.org/abs/gr-qc/9906123}
  {gr-qc/9906123} \BibitemShut {NoStop}%
\bibitem [{\citenamefont {{Ashby}}\ and\ \citenamefont
  {{Bertotti}}(2010)}]{ashby:2010cg}%
  \BibitemOpen
  \bibfield  {author} {\bibinfo {author} {\bibfnamefont {N.}~\bibnamefont
  {{Ashby}}}\ and\ \bibinfo {author} {\bibfnamefont {B.}~\bibnamefont
  {{Bertotti}}},\ }\bibfield  {title} {\enquote {\bibinfo {title} {{Accurate
  light-time correction due to a gravitating mass}},}\ }\href {\doibase
  10.1088/0264-9381/27/14/145013} {\bibfield  {journal} {\bibinfo  {journal}
  {Classical Quantum Gravity}\ }\textbf {\bibinfo {volume} {27}},\ \bibinfo
  {eid} {145013} (\bibinfo {year} {2010})},\ \Eprint
  {http://arxiv.org/abs/0912.2705} {arXiv:0912.2705 [gr-qc]} \BibitemShut
  {NoStop}%
\bibitem [{\citenamefont {{Linet}}\ and\ \citenamefont
  {{Teyssandier}}(2016)}]{linet:2016pr}%
  \BibitemOpen
  \bibfield  {author} {\bibinfo {author} {\bibfnamefont {B.}~\bibnamefont
  {{Linet}}}\ and\ \bibinfo {author} {\bibfnamefont {P.}~\bibnamefont
  {{Teyssandier}}},\ }\bibfield  {title} {\enquote {\bibinfo {title} {{Time
  transfer functions in Schwarzschild-like metrics in the weak-field limit: A
  unified description of Shapiro and lensing effects}},}\ }\href {\doibase
  10.1103/PhysRevD.93.044028} {\bibfield  {journal} {\bibinfo  {journal}
  {\prd}\ }\textbf {\bibinfo {volume} {93}},\ \bibinfo {eid} {044028} (\bibinfo
  {year} {2016})},\ \Eprint {http://arxiv.org/abs/1511.04284} {arXiv:1511.04284
  [gr-qc]} \BibitemShut {NoStop}%
\bibitem [{\citenamefont {{Tempel}}\ \emph {et~al.}(2016)\citenamefont
  {{Tempel}}, \citenamefont {{Kipper}}, \citenamefont {{Tamm}}, \citenamefont
  {{Gramann}}, \citenamefont {{Einasto}}, \citenamefont {{Sepp}},\ and\
  \citenamefont {{Tuvikene}}}]{tempel:2016aa}%
  \BibitemOpen
  \bibfield  {author} {\bibinfo {author} {\bibfnamefont {E.}~\bibnamefont
  {{Tempel}}}, \bibinfo {author} {\bibfnamefont {R.}~\bibnamefont {{Kipper}}},
  \bibinfo {author} {\bibfnamefont {A.}~\bibnamefont {{Tamm}}}, \bibinfo
  {author} {\bibfnamefont {M.}~\bibnamefont {{Gramann}}}, \bibinfo {author}
  {\bibfnamefont {M.}~\bibnamefont {{Einasto}}}, \bibinfo {author}
  {\bibfnamefont {T.}~\bibnamefont {{Sepp}}}, \ and\ \bibinfo {author}
  {\bibfnamefont {T.}~\bibnamefont {{Tuvikene}}},\ }\bibfield  {title}
  {\enquote {\bibinfo {title} {{Friends-of-friends galaxy group finder with
  membership refinement. Application to the local Universe}},}\ }\href
  {\doibase 10.1051/0004-6361/201527755} {\bibfield  {journal} {\bibinfo
  {journal} {Astron. Astrophys.}\ }\textbf {\bibinfo {volume} {588}},\ \bibinfo
  {eid} {A14} (\bibinfo {year} {2016})},\ \Eprint
  {http://arxiv.org/abs/1601.01117} {arXiv:1601.01117} \BibitemShut {NoStop}%
\bibitem [{\citenamefont {{Tully}}(2015)}]{tully:2015aj}%
  \BibitemOpen
  \bibfield  {author} {\bibinfo {author} {\bibfnamefont {R.~B.}\ \bibnamefont
  {{Tully}}},\ }\bibfield  {title} {\enquote {\bibinfo {title} {{Galaxy Groups:
  A 2MASS Catalog}},}\ }\href {\doibase 10.1088/0004-6256/149/5/171} {\bibfield
   {journal} {\bibinfo  {journal} {Astron. J.}\ }\textbf {\bibinfo {volume}
  {149}},\ \bibinfo {eid} {171} (\bibinfo {year} {2015})},\ \Eprint
  {http://arxiv.org/abs/1503.03134} {arXiv:1503.03134} \BibitemShut {NoStop}%
\bibitem [{cod()}]{code}%
  \BibitemOpen
  \href@noop {} {}\bibinfo {howpublished} {Code to compute Shapiro delay with
  galaxy catalogs,
  \url{https://git.ligo.org/olivier.minazzoli/shapiro_cosmo_shortcomings_public}}\BibitemShut
  {NoStop}%
\bibitem [{\citenamefont {Coulter}\ \emph {et~al.}(2017)\citenamefont {Coulter}
  \emph {et~al.}}]{Coulter:2017wya}%
  \BibitemOpen
  \bibfield  {author} {\bibinfo {author} {\bibfnamefont {D.~A.}\ \bibnamefont
  {Coulter}} \emph {et~al.},\ }\bibfield  {title} {\enquote {\bibinfo {title}
  {{Swope Supernova Survey 2017a (SSS17a), the Optical Counterpart to a
  Gravitational Wave Source}},}\ }\href {\doibase 10.1126/science.aap9811}
  {\bibfield  {journal} {\bibinfo  {journal} {Science}\ }\textbf {\bibinfo
  {volume} {358}},\ \bibinfo {pages} {1556} (\bibinfo {year} {2017})},\ \Eprint
  {http://arxiv.org/abs/1710.05452} {arXiv:1710.05452 [astro-ph.HE]}
  \BibitemShut {NoStop}%
\bibitem [{\citenamefont {Hoffman}\ \emph {et~al.}(2017)\citenamefont
  {Hoffman}, \citenamefont {Pomar{\`e}de}, \citenamefont {Tully},\ and\
  \citenamefont {Courtois}}]{Hoffman:2017ako}%
  \BibitemOpen
  \bibfield  {author} {\bibinfo {author} {\bibfnamefont {Y.}~\bibnamefont
  {Hoffman}}, \bibinfo {author} {\bibfnamefont {D.}~\bibnamefont
  {Pomar{\`e}de}}, \bibinfo {author} {\bibfnamefont {R.~B.}\ \bibnamefont
  {Tully}}, \ and\ \bibinfo {author} {\bibfnamefont {H.}~\bibnamefont
  {Courtois}},\ }\bibfield  {title} {\enquote {\bibinfo {title} {{The Dipole
  Repeller}},}\ }\href {\doibase 10.1038/s41550-016-0036} {\bibfield  {journal}
  {\bibinfo  {journal} {Nat. Astron.}\ }\textbf {\bibinfo {volume} {1}},\
  \bibinfo {pages} {0036} (\bibinfo {year} {2017})},\ \Eprint
  {http://arxiv.org/abs/1702.02483} {arXiv:1702.02483 [astro-ph.CO]}
  \BibitemShut {NoStop}%
\bibitem [{Viz()}]{Vizier}%
  \BibitemOpen
  \href@noop {} {}\bibinfo {howpublished} {VizieR Information System,
  \href{https://doi.org/10.26093/cds/vizier}{https://doi.org/10.26093/cds/vizier}}\BibitemShut
  {NoStop}%
\bibitem [{\citenamefont {Ochsenbein}\ \emph {et~al.}(2000)\citenamefont
  {Ochsenbein}, \citenamefont {Bauer},\ and\ \citenamefont
  {Marcout}}]{Ochsenbein:2000th}%
  \BibitemOpen
  \bibfield  {author} {\bibinfo {author} {\bibfnamefont {F.}~\bibnamefont
  {Ochsenbein}}, \bibinfo {author} {\bibfnamefont {P.}~\bibnamefont {Bauer}}, \
  and\ \bibinfo {author} {\bibfnamefont {J.}~\bibnamefont {Marcout}},\
  }\bibfield  {title} {\enquote {\bibinfo {title} {{The VizieR database of
  astronomical catalogues}},}\ }\href {\doibase 10.1051/aas:2000169} {\bibfield
   {journal} {\bibinfo  {journal} {Astron. Astrophys. Suppl. Ser.}\ }\textbf
  {\bibinfo {volume} {143}},\ \bibinfo {pages} {23} (\bibinfo {year} {2000})},\
  \Eprint {http://arxiv.org/abs/astro-ph/0002122} {arXiv:astro-ph/0002122
  [astro-ph]} \BibitemShut {NoStop}%
\bibitem [{\citenamefont {{Poisson}}\ and\ \citenamefont
  {{Will}}(2014)}]{2014grav.book.....P}%
  \BibitemOpen
  \bibfield  {author} {\bibinfo {author} {\bibfnamefont {E.}~\bibnamefont
  {{Poisson}}}\ and\ \bibinfo {author} {\bibfnamefont {C.~M.}\ \bibnamefont
  {{Will}}},\ }\href@noop {} {\emph {\bibinfo {title} {{Gravity}}}}\ (\bibinfo
  {publisher} {Cambridge University Press},\ \bibinfo {address} {Cambridge,
  UK},\ \bibinfo {year} {2014})\BibitemShut {NoStop}%
\bibitem [{\citenamefont {{Borwein}}\ \emph {et~al.}(1985)\citenamefont
  {{Borwein}}, \citenamefont {{Borwein}},\ and\ \citenamefont
  {{Taylor}}}]{1985JMP....26.2999B}%
  \BibitemOpen
  \bibfield  {author} {\bibinfo {author} {\bibfnamefont {D.}~\bibnamefont
  {{Borwein}}}, \bibinfo {author} {\bibfnamefont {J.~M.}\ \bibnamefont
  {{Borwein}}}, \ and\ \bibinfo {author} {\bibfnamefont {K.~F.}\ \bibnamefont
  {{Taylor}}},\ }\bibfield  {title} {\enquote {\bibinfo {title} {{Convergence
  of lattice sums and Madelung's constant}},}\ }\href {\doibase
  10.1063/1.526675} {\bibfield  {journal} {\bibinfo  {journal} {J. Math.
  Phys.}\ }\textbf {\bibinfo {volume} {26}},\ \bibinfo {pages} {2999} (\bibinfo
  {year} {1985})}\BibitemShut {NoStop}%
\end{thebibliography}
\end{document}